\documentclass[reprint,showpacs,preprintnumbers,amsmath,amssymb,aps,prd,nofootinbib,floatfix,longbibliography]{revtex4-2}

\usepackage{bbold}
\usepackage{hyperref}

\usepackage{graphicx}

\usepackage{color}

\newcommand{\comment}[1]{#1}

\newcommand{\highlight}[1]{#1}
\newcommand{\appsection}[1]{\section{\uppercase{#1}}}

\begin{document}

\title{Atom interferometer as a freely falling clock for time-dilation measurements}

\author{Albert Roura}
\affiliation{German Aerospace Center (DLR), Institute of Quantum Technologies,
Wilhelm-Runge-Stra\ss e 10, 89081 Ulm, Germany}


\begin{abstract}
Light-pulse atom interferometers based on single-photon transitions are a promising tool for gravitational-wave detection in the mid-frequency band and the search for ultralight dark-matter fields.
Here we present a novel measurement scheme that enables their use as freely falling clocks directly measuring relativistic time-dilation effects.
The proposal is particularly timely because it can be implemented with no additional requirements in Fermilab's MAGIS-100 experiment or even in the 10-m prototypes that are expected to start operating very soon.
This will allow the unprecedented measurement of gravitational time dilation in a local experiment with freely falling atoms, which is out of reach even for the best atomic-fountain clocks based on microwave transitions.
The results are supported by a comprehensive treatment of relativistic effects in this kind of interferometers as well as a detailed analysis of the main systematic effects.
Furthermore, the theoretical methods developed here constitute a valuable tool for modelling light-pulse atom interferometers based on single-photon transitions in general.
\end{abstract}

\pacs{}

\maketitle

\section{Introduction}
\label{sec:introduction}

The great potential of matter-wave interferometers for high-precision inertial sensing was recognized early on \cite{clauser88,borde89} and since the first experimental realizations three decades ago \cite{carnal91,keith91,riehle91,kasevich91} atom interferometric quantum sensors have become an essential tool for both fundamental research and 
practical applications \cite{bongs19}.
Indeed, besides their use in gravimetry \cite{peters99,merlet10,stray22} and as highly accurate gyrometers \cite{savoie18,gautier22}, light-pulse atom interferometers relying on Raman or Bragg diffraction have proven to be very valuable for a wide range of applications in fundamental physics, including precise measurements of the fine-structure constant \cite{parker18,morel20} and the gravitational constant \cite{rosi14}, tests of the universality of free fall (UFF) \cite{asenbaum20,schlippert14,rosi17b}, searches for dark-energy candidates \cite{hamilton15b,jaffe17,sabulsky19} and the measurement of spacetime curvature effects on delocalized quantum superpositions \cite{asenbaum17,overstreet22,roura22}.

More recently, a new kind of atom interferometers \cite{graham13} based on single-photon transitions between the two clock states in atoms such as Sr or Yb, which are commonly employed in optical atomic clocks, has been receiving increasing attention \cite{hu17,hu20,rudolph20,wilkason22}. A key appealing feature is the possibility of having single-baseline gravitational-wave detectors \cite{yu11,graham13} that are immune to laser phase noise (in contrast to two or more baselines needed for optical interferometers or for schemes employing atom interferometers based on two-photon transitions \cite{canuel20a}) and can also be exploited 
to search for ultralight dark matter~\cite{arvanitaki18,gue24}. 
Such detectors involve a gradiometer-type configuration consisting of two spatially separated atom interferometers interrogated by a common laser beam and their sensitivity is proportional to the length of the baseline between the two interferometers. Thus, although ultimate sensitivities could be reached in space \cite{hogan16,graham17,el-neaj20,tino19}, where baselines of thousands or even millions of kilometers are possible and there is a gravitationally quieter environment,
kilometer-scale detectors on ground are also being considered \cite{abe21,abend23}.
As an intermediate step, a 100-m atomic fountain prototype, MAGIS-100 \cite{abe21,magis}, is currently being assembled at Fermilab and a similar set-up is under study at CERN \cite{arduini23}. Furthermore, closely related efforts are also being pursued in the UK~\cite{badurina20} and China~\cite{zhan19}.

While the 100-m prototypes will play a crucial role for technology development and proof-of-principle experiments, it is expected that the attainable sensitivities will be insufficient for gravitational-wave detection. Similarly, the likely outcome of the search for ultralight dark-matter fields may simply be an
improvement of the bounds
for the couplings to the Standard Model sector.
It is therefore particularly important to devise experiments that go beyond mere null tests and enable the actual measurement 
of interesting fundamental physics effects within reach of the planned detector sensitivities.
With this spirit in mind, we will show here that such long-baseline 
facilities 
offer the opportunity to perform unprecedented measurements of relativistic effects with freely falling atoms. Indeed, thanks to the measurement scheme proposed \comment{below}, an atom interferometer can be employed as a freely falling clock for time-dilation measurements capable of outperforming state-of-the-art atomic fountain clocks by several orders of magnitude.

These effects, which include both special relativistic and gravitational time dilation, have been measured with clocks on rockets \cite{vessot80}, satellites \cite{delva18,hermann18} or planes \cite{hafele72} that are compared to ground stations. The gravitational redshift has also been measured by comparing static atomic clocks at different heights \cite{takamoto20}, but not \comment{in a local comparison} to a clock involving freely falling atoms.
A natural possibility in this respect would be to consider atomic fountain clocks with cold atoms 
\cite{sullivan01}. However, the best precisions achieved with such atomic clocks, which rely on a microwave transition, fall short%
\footnote{\highlight{Sensitivity to gravitational and special relativistic time dilation is possible when comparing two atomic-fountain clocks at sufficiently different heights or latitudes, but not with a single atomic fountain compared to a static clock at the same location.}}
by \comment{an} order of magnitude \cite{roura20a,ashby21}.
In order to reach higher sensitivities, one could alternatively use 
the same kind of atoms and clock transitions employed in optical atomic clocks \cite{poli14,ludlow15}.
Nevertheless, implementing this idea is not straightforward because contrary to atoms trapped in an optical lattice, recoil effects are non-negligible for freely falling atoms and one is naturally led to consider light-pulse atom interferometers based on single-photon transitions.
Moreover, identifying time-dilation effects in such interferometers poses a number of conceptual and practical challenges.

In this \comment{article} we will show how those challenges can be overcome 
and will propose a measurement scheme that can be experimentally realized in a facility such as MAGIS-100 without any additional requirements.
Furthermore, the theoretical methods developed here will be very valuable for a detailed modelling of light-pulse atom interferometers based on single-photon transitions in general.
\highlight{The technical details of the theoretical treatment, the phase-shift calculation and the analysis of the main systematic effects are provided in eight appendices.
In addition, the effects of violations of the equivalence principle are considered in Appendix~\ref{app:violations}.}

\section{Relativistic effects in freely falling clocks}
\label{sec:ff_clocks}

The proper time along a world line $X^\mu(\lambda)$ is an invariant quantity 
that generalizes to curved spacetimes the notion of length along an arbitrary curve in Riemannian geometry, and it corresponds to the time that an ideal clock following that world line would measure.
For non-relativistic velocities and weak gravitational fields one can consider a post-Newtonian expansion in powers of $1/c^2$. Up to first order the proper time is given by
\begin{equation}
\Delta \tau = \int^{t}_{t_0} dt' \left( 1 - \frac{1}{2\, c^2}\! \left( \frac{d \mathbf{X}}{d t'} \right)^2
+\, \frac{1}{c^2}\, U (t',\mathbf{X}) \right)
\label{eq:proper_time} ,
\end{equation}
where the parametrization $X^\mu(t') = \big(c\, t', \mathbf{X}(t') \big)$ in terms of the usual coordinates in a post-Newtonian expansion~\cite{misner73} has been employed. Here $U (t',\mathbf{X})$ corresponds to the Newtonian gravitational potential and for the particular case of a uniform gravitational field it reduces to $U (t',\mathbf{x}) = U_0 - \mathbf{g} \cdot (\mathbf{x} - \mathbf{x}_0)$.

Atomic clocks rely on the transition between the electronic ground state $| \mathrm{g} \rangle$ and a sufficiently long-lived excited state $| \mathrm{e} \rangle$ with an energy difference $\Delta E$. If we consider a quantum superposition of these two internal states, commonly known as the \emph{clock states}, the relative phase between them will grow with time and 
\comment{is directly related to} the elapsed proper time $\Delta \tau$:
\begin{equation}
\big| \Phi (\tau) \big\rangle \propto \frac{1}{\sqrt{2}}
\left( | \mathrm{g} \rangle + e^{- i \Delta E \, \Delta\tau / \hbar}\, \comment{ e^{i \varphi_\mathrm{i}} }
| \mathrm{e} \rangle \right)
\label{eq:clock_evolution} ,
\end{equation}
where the left- and right-hand sides are equal up to a global phase factor and we have included a possible phase $\varphi_\mathrm{i}$ associated with the clock initialization.
In a detailed description one also needs to consider the quantum state of the atom's center-of-mass
degree of freedom. Nevertheless, as shown in Ref.~\cite{roura20a} and briefly reviewed in \comment{Appendix~\ref{app:propagation}}, the evolution of the atomic wave packets 
can be conveniently formulated in terms of central trajectories and centered wave packets.
In fact, for our considerations below 
it will be sufficient to focus on the central trajectories, which satisfy the classical equations of motion,
and the proper time calculated along their corresponding world lines.

During its free evolution between laser pulses an atomic wave packet acquires
a propagation phase $\exp(i\, \mathcal{S}_n / \hbar)$ with
\begin{equation}
\mathcal{S}_n = - m_n c^2 \int^{\tau}_{\tau_0} d\tau'
\label{eq:prop_phase} .
\end{equation}
where the subindex $n=1,2$ labels the internal state, and the rest masses for the ground and excited states are given by $m_1 = m$ and $m_2 = m + \Delta E / c^2$ respectively.
For non-relativistic velocities and weak gravitational fields one can employ Eq.~\eqref{eq:proper_time} when calculating the proper time, and the propagation phase reduces then to the classical action plus a rest-mass energy term.

\begin{figure}[ht]
\begin{center}
\includegraphics[width=7.5cm]{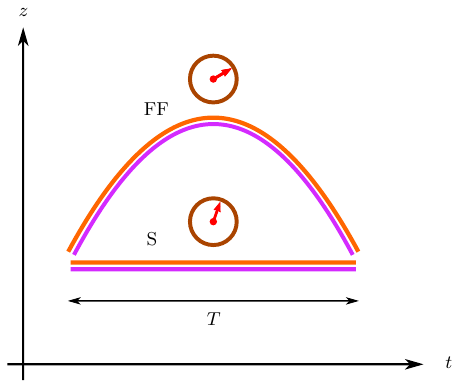}
\end{center}
\caption{Comparison between static (S) and freely falling (FF) atoms acting as quantum clocks and involving a quantum superposition of two different internal states. The accumulated relative phase between the ground (purple) and excited (orange) state is proportional to the elapsed proper time, which differs in both cases due to relativistic time-dilation effects.
The curves in this spacetime diagram correspond to the central trajectories of the atomic wave packets in the laboratory frame.
}
\label{fig:clock_comparison}
\end{figure}

As an example, let us consider an atom in a superposition of the internal states $| \mathrm{g} \rangle$ and $| \mathrm{e} \rangle$ freely falling in a uniform gravitational field. The central trajectory of the atomic wave packets for both internal states corresponds to the parabolic world line depicted in Fig.~\ref{fig:clock_comparison},
and the difference between their propagation phases leads to a relative phase $\delta\phi = (\mathcal{S}_2 - \mathcal{S}_1) / \hbar = - (\Delta E / \hbar)\, \Delta\tau$.
Evaluating Eq.~\eqref{eq:proper_time} for this case, one obtains
\begin{equation}
\delta\phi = - (\Delta E / \hbar) \left( \left( 1 + U_0 / c^2 \right) T + \frac{1}{24} \frac{g^2 T^3}{c^2} \right)
\label{eq:ff-clock_phase0} ,
\end{equation}
\comment{where $\Delta t = T$ is the time-coordinate difference between the 
intersections of both world lines.}
In contrast, for an atom trapped in a suitable potential so that the central position of the atomic wave packets remains at constant height in the laboratory frame, one would obtain the result in Eq.~\eqref{eq:ff-clock_phase0} but without the last term \cite{roura20a}. 
Hence, it is precisely this term that corresponds to the difference between the proper times measured by a freely falling clock and a static one in a uniform field. 
A natural way of implementing such a measurement would be to compare an atomic fountain clock employing Rb or Cs atoms, and relying on the microwave transition between the two hyperfine ground states, with an optical atomic clock involving Sr or Yb atoms trapped in an optical lattice. However, the effect is \comment{an} order of magnitude smaller than the best accuracy achieved by atomic fountain clocks \cite{roura20a,ashby21}.

Before we explore the possibility of using instead Sr or Yb as freely falling atoms, it is helpful to 
analyze first the \comment{situation} displayed in Fig.~\ref{fig:ideal_clock}, assuming for the moment an ideal clock. Contrary to the case of Fig.~\ref{fig:clock_comparison}, the heights at the initial and final times are not necessarily the same and the internal states are swapped at some intermediate time. Moreover, synchronization becomes a non-trivial 
issue (even conceptually) due to the relativity of simultaneity for spatially separated events.
Thus, we will consider simultaneity hypersurfaces with respect to the laboratory frame for the initialization,
internal-state inversion 
and final read-out as well as time differences $\Delta t = T$ between them, as indicated in Fig.~\ref{fig:ideal_clock}.
Proceeding as we did above to calculate the difference of propagation phases for the two internal states by evaluating Eq.~\eqref{eq:proper_time} along the central trajectory $\mathbf{X} (t)$ and taking into account the state inversion at the intermediate time, one gets the following result for the relative phase:
\begin{equation}
\delta\phi = - 2 \,(\Delta E / \hbar) \left( \mathbf{v}_0 \cdot \mathbf{g}\, T^2 + g^2 T^3  \right) / c^2
\label{eq:clock_phase} ,
\end{equation}
where $\mathbf{v}_0 = \left. (d \mathbf{X} / dt) \right|_{t_0}$ 
and $T$ can be regarded here as the proper time measured by a static clock at constant height because the corrections proportional to $U_0 / c^2$ would give rise to terms of higher order in $1/c^2$.
In this case the result is independent of $\mathbf{X}_0$, in contrast with the right-hand side of Eq.~\eqref{eq:ff-clock_phase0}, which depends implicitly on $\mathbf{X}_0$ and $\mathbf{v}_0$ through its dependence on $U_0$ and the particular choice $\mathbf{v}_0 = \mathbf{g}\, T / 2$ that was made.


For an actual implementation of the measurement depicted in Fig.~\ref{fig:ideal_clock} one could contemplate using the Doppler-free two-photon transition investigated in Ref.~\cite{alden14}, which would guarantee a vanishing momentum transfer and simultaneity in the laboratory frame \cite{roura20a}.
However, pulses relying on such a transition require high laser power and a special set-up. Moreover, despite the vanishing momentum transfer, the atoms experience a residual recoil that depends on their velocity when the pulse is applied~\cite{roura20a}. The resulting modification of the central trajectory is rather small, but leads to a spurious phase-shift contribution comparable to the time dilation effect that we are interested in.
Instead, with a suitable measurement scheme a light-pulse atom interferometer based on single-photon transitions, \highlight{where the atomic wave packets are split, redirected and recombined by the laser pulses,}
can be employed to measure such relativistic effects. 
Indeed, the scheme presented in the next section is equivalent to the ideal freely falling clock in Fig.~\ref{fig:ideal_clock}, does not suffer from the drawbacks of the Doppler-free transition and can be experimentally implemented without additional requirements to those already planned for facilities such as MAGIS-100.

\section{Atom interferometer acting as a freely falling clock}
\label{sec:atom_interferometer}

\begin{figure*}[t]
\onecolumngrid
\begin{center}
\includegraphics[width=8.0cm]{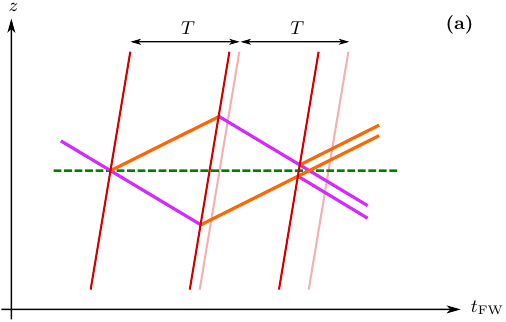}
\hspace{6.0ex}
\includegraphics[width=8.0cm]{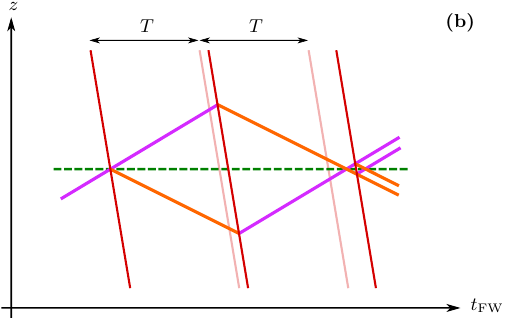}
\end{center}
\caption{Spacetime diagram of the atom interferometer in the freely falling frame associated with the mid-point trajectory (dashed line). The color of the central trajectories for the atomic wave packets propagating along the two interferometer arms depends on the internal state, whether ground (purple) or excited (orange). 
In the absence of Doppler shift and time dilation the central wave fronts of the laser pulses (red) are equally separated (pale red). Any uncompensated Doppler factor that remains after the frequency chirp leads to opposite shifts of the laser wave fronts in the two reversed interferometers (a and b). In contrast, smaller time dilation effects lead to the same shift in both interferometers. \comment{The shifts of the laser wave fronts have been exaggerated in both diagrams for illustrative purposes.}}
\label{fig:ff_frame}
\end{figure*}

In order to study 
atom interferometers based on single-photon transitions and their possible use as freely falling clocks, it is particularly convenient to consider the freely falling frame associated with the mid-point trajectory between the two interferometer arms \comment{(Fermi-Walker frame).} In such a frame the spacetime coordinates of the mid-point world line take the simple form $\bar{X}^\mu(t_\text{FW}) = \big(c\, t_\text{FW}, \mathbf{0} \big)$ and the comoving time coordinate $t_\text{FW}$ coincides with the proper time  $\bar{\tau}$ along the world line. 
Furthermore, in this frame the spacetime trajectories for light rays correspond to simple straight lines, except for small curvature effects that are completely negligible in our case (see Appendix~\ref{sec:non-uniform_field}). Hence, \comment{while describing gravitational effects on light propagation as well as effects due to the motion of the atomic wave packets can be more involved in the laboratory frame,} in the freely falling frame they simply amount to shifts of straight lines with fixed slope,
as illustrated in Fig.~\ref{fig:ff_frame}.

\begin{figure}[t]
\begin{center}
\includegraphics[width=7.0cm]{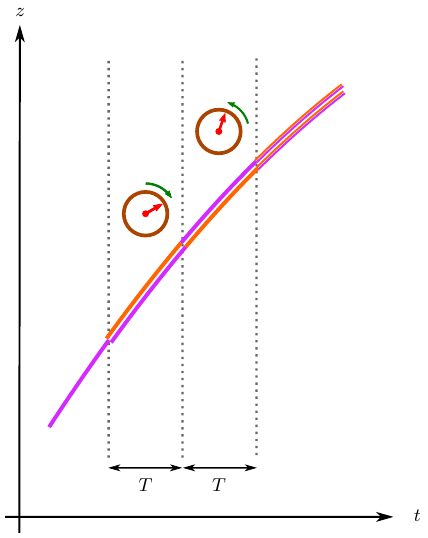}
\end{center}
\caption{Spacetime diagram for freely falling atoms in a quantum superposition of the ground (purple) and excited (orange) state that undergo an internal-state inversion at the intermediate time.
\highlight{The atoms act as a clock where proper time is encoded in the relative phase between the two components of the superposition and that runs backwards after the inversion.} 
A~static clock serves as a time reference and simultaneity hypersurfaces in the laboratory frame (dotted lines) are considered for the initialization, inversion and read-out events. Despite equal-time separation $T$ in the reference clock, time dilation effects lead to an imbalance between the proper times before and after state inversion for the freely falling atoms.}
\label{fig:ideal_clock}
\end{figure}

In fact, these shifts can be calculated by considering the intersection of the light rays with the world line of the mid-point trajectory from the point of view of the laboratory frame, and can be naturally interpreted in terms of Doppler shift and relativistic time dilation%
\footnote{For convenience, throughout this paper the term ``Doppler shift'' will refer exclusively to the retardation effects due to the finite speed of light and the motion of the atoms. This differs from the usual terminology in \highlight{special} relativistic treatments, which also includes \highlight{the contribution of time dilation.}}.
The details can be found in Appendix~\ref{app:ff_frame}, 
but the key aspects can be summarized as follows.
For a \comment{stationary} spacetime a natural choice of time coordinate in the laboratory frame is the time $t$ associated with the time-translation symmetry 
of the spacetime metric. If we denote by $\bar{t}$ the time at which a given light ray intersects the mid-point trajectory, the time separation $dt$ between two light rays emitted from a fixed position in the laboratory frame and the time difference $d\bar{t}$ between their intersections with the mid-point trajectory are related by
\begin{equation}
\quad \ 
\frac{d \bar{t}}{d t} =  \left( \frac{1}{1 - \hat{\mathbf{n}} \cdot \bar{\mathbf{v}} / c} \right)
\label{eq:Doppler} ,
\end{equation}
which corresponds to the Doppler shift and where $\hat{\mathbf{n}}$ is the direction of the light rays and $\bar{\mathbf{v}} = d \bar{\mathbf{X}} / d\bar{t}$ is the velocity of the mid-point trajectory in the laboratory frame.
On the other hand, the relation between the time separation $d\bar{t}$ in the laboratory frame and the proper time $d\bar{\tau}$ elapsed along the mid-point world line between the two light-ray intersections is given by
\begin{equation}
\frac{d \bar{\tau}}{d \bar{t}} \,=\, 1 - \frac{1}{2\, c^2}\! \left( \frac{d \bar{\mathbf{X}}}{d \bar{t}} \right)^2
+\, \frac{1}{c^2}\, U \big( \bar{t},\bar{\mathbf{X}} \big)
\label{eq:time_dilation} ,
\end{equation}
where the second and third terms on the right-hand side correspond, respectively, to special relativistic and gravitational time dilation, and terms of order $1/c^4$ or higher have been neglected.

\highlight{As we will see,} the phase shift $\delta\phi$ between the two arms of the atom interferometer can be inferred from the positions of the laser wave fronts in the freely falling frame and each wave front (regarded as a null hypersurface in spacetime) can be characterized by its phase $\varphi$, which is a frame-independent quantity.
In particular, one has the following relation between the phase $\varphi$ and the proper time $\bar{\tau}$ at which the wave front intersects the mid-point world line:
\begin{equation}
\quad \ 
\frac{d \bar{\tau}}{d \varphi} = \frac{d \bar{\tau}}{d \bar{t}} \, \frac{d \bar{t}}{d t} \left( \frac{d t}{d \varphi} \right)
=  \frac{d \bar{\tau}}{d \bar{t}} \left( \frac{1}{1 - \hat{\mathbf{n}} \cdot \bar{\mathbf{v}} / c} \right)
\left( \frac{d t}{d \varphi} \right)
\label{eq:proper-time_phase} ,
\end{equation}
where $d \bar{\tau} / d \bar{t}$ is given by Eq.~\eqref{eq:time_dilation} and $(d t / d \varphi) = 1 / \omega$ corresponds to the inverse of the (possibly time-dependent) angular frequency with which the electromagnetic wave is emitted from a fixed position in the laboratory frame.
From Eq.~\eqref{eq:proper-time_phase} it is clear that one can, in principle, compensate the Doppler shift through a suitable frequency chirp of the emitted radiation, 
\begin{equation}
\left( \frac{d t}{d \varphi} \right)_\text{chirp}
\!=\, \Big( 1 - \hat{\mathbf{n}} \cdot \bar{\mathbf{v}}' / c \Big) \left( \frac{d t}{d \varphi} \right)_0
\label{eq:chirp1} ,
\end{equation}
provided that $\bar{\mathbf{v}}'(\bar{t}) = \bar{\mathbf{v}}(\bar{t})$
and where $(d t / d \varphi)_0 = 1 / \omega_0$ corresponds to the inverse of the unchirped frequency. In practice, however, a perfect match will not be possible. 
Hence, \comment{for a mid-point trajectory 
\begin{equation}
\bar{\mathbf{X}}(\bar{t}) = \bar{\mathbf{X}}_0 +  \bar{\mathbf{v}}_0 \left( \bar{t} - \bar{t}_0 \right)
+ \frac{1}{2}\, \mathbf{g} \left( \bar{t} - \bar{t}_0 \right)^2
\label{eq:mid-point1} ,
\end{equation}
}%
we will 
actually have $\bar{\mathbf{v}}' (\bar{t}) = \bar{\mathbf{v}}'_0 + \mathbf{g}' \left( \bar{t} - \bar{t}_0 \right)$ with small non-vanishing deviations $\Delta\bar{\mathbf{v}}_0 = \bar{\mathbf{v}}_0 - \bar{\mathbf{v}}'_0$ and $\Delta\mathbf{g} = \mathbf{g} - \mathbf{g}'$.
\linebreak[4]
 \comment{Eq.~\eqref{eq:chirp1} determines the frequency chirp that must be applied to the laser carrier wave, but it also implies a slight change of the central time of the pulse envelope. Pulse timings cannot be controlled so precisely as the phase of the carrier wave, but the interferometer signal is much less sensitive to imperfections in the pulse timing, as discussed in Appendix~\ref{sec:pulse_timings}.}

%
%


\twocolumngrid

 \highlight{
A more detailed derivation of the phase shift $\delta\phi$ can be found in Appendix~\ref{app:phase-shift}, but the key idea can be intuitively understood from Fig.~\ref{fig:ff_frame}. For the case of perfect compensation of the Doppler factor and in the absence of time-dilation effects, the diagram is symmetric with respect to the intersection point 
of the mid-point world line and the central wave front of the second laser pulse. The time spent in the excited state is therefore the same along the two interferometer arms and $\delta\phi$ vanishes. On the other hand, time dilation leads to a small shift of the wave fronts and their intersection with the mid-point trajectory that can be calculated by integrating Eq.~\eqref{eq:time_dilation} with respect to $\bar{t}$ ---after substitution of Eq.~\eqref{eq:mid-point1}--- and is of order $1/c^2$. The resulting change of the propagation phases can then be directly obtained from the calculation of the proper time along the mid-point trajectory because additional contributions to the actual proper time along the central trajectories are further suppressed by additional powers of $v_\text{rec} / c$ \comment{as shown in Appendix~\ref{app:arm_trajectories}, where the recoil velocity $\mathbf{v}_\text{rec}$ is more precisely defined.}
The computation of the proper time along the mid-point trajectory, which is equivalent to 
the ideal clock in Fig.~\ref{fig:ideal_clock}, leads to the main contribution in the full result for $\delta\phi$:
}
\begin{equation}
\delta\phi = - 2 \,(\Delta E / \hbar) \left( \bar{\mathbf{v}}_0 \cdot \mathbf{g}\, T^2 + g^2 T^3  \right) / c^2
+ \delta\phi_\text{corr} \,
\label{eq:phase_shift} ,
\end{equation}
where terms suppressed by higher powers of $1/c$ have been neglected and $\delta\phi_\text{corr}$, which accounts for any imperfect matching of the chirp rate, is given by
\comment{
\begin{align}
\delta\phi_\text{corr} &= \frac{\Delta E}{\hbar} \left[ \frac{\hat{\mathbf{n}} \cdot \Delta\mathbf{g}}{c}\, T^2
+ 2\, \frac{\Delta\bar{\mathbf{v}}_0 \cdot \mathbf{g}}{c^2}\, T^2
\right. \nonumber \\
& \qquad\qquad \left. +\, \frac{\bar{\mathbf{v}}_0 \cdot \Delta\mathbf{g}}{c^2}\, T^2
+ 3\, \frac{ \mathbf{g} \cdot \Delta\mathbf{g}}{c^2}\, T^3
\right]
\label{eq:corrections1} .
\end{align}
}%
Here terms involving higher powers of $1/c$ or higher orders in $\Delta\bar{\mathbf{v}}_0$ and $\Delta\mathbf{g}$ have been omitted, and we have assumed for simplicity that $\hat{\mathbf{n}}$, $\bar{\mathbf{v}}_0$ and $\mathbf{g}$ are all aligned. The expression for the general case, \comment{corresponding to Eq.~\eqref{eq:corrections4}, 
can be found in Appendix~\ref{app:phase-shift}.}

\highlight{Note that for perfect pulse timings a time-independent shift of the laser frequency does not modify $\delta\phi$ to first order in the frequency shift. The same applies to a small change of $\bar{\mathbf{v}}'_0$ in Eq.~\eqref{eq:chirp1}, 
which only contributes to $\delta\phi_\text{corr}$ through a higher-order term suppressed by an additional factor $g\, T / c$. Such a milder sensitivity to time-independent frequency shifts relaxes the requirements on accuracy and long-term stability of the laser frequency.
These requirements are further relaxed for the gradiometric configuration considered below.
In the presence of pulse timing errors, on the other hand, there is a trade-off between the requirements on the frequency detuning $\delta$ and the pulse-timing shift $\Delta T$, as discussed in Appendix~\ref{sec:pulse_timings}.
}

\subsection*{Mitigation of spurious contributions}

Comparing Eqs.~\eqref{eq:clock_phase} and \eqref{eq:phase_shift}, we can see that the result for the atom interferometer coincides with that for an ``ideal'' clock following the mid-point trajectory as long as the phase-shift correction $\delta\phi_\text{corr}$ can be neglected.
It is therefore important to analyze the various contributions to $\delta\phi_\text{corr}$.
The first term on the right-hand side of Eq.~\eqref{eq:corrections1} is of order $1/c$ and can be substantially larger than the signal that we are interested in, namely the two terms of order $1/c^2$ explicitly written in Eq.~\eqref{eq:phase_shift}. Fortunately, the contribution of this term can be further suppressed by considering also a \emph{reversed interferometer} with the same mid-point trajectory but $\hat{\mathbf{n}} \to - \hat{\mathbf{n}}$, i.e.\ opposite propagation direction for the laser pulses. 
When considering the \comment{semisum} of the phase shifts obtained for both interferometers,
the first term on the right-hand side of Eq.~\eqref{eq:corrections1}, which is linear in $\hat{\mathbf{n}}$, will cancel out, whereas the terms of interest in Eq.~\eqref{eq:phase_shift}, which are independent of $\hat{\mathbf{n}}$, will remain unchanged.
Similarly, the leading phase-shift corrections due to gravity gradients and rotations,  \comment{discussed in Appendix~\ref{app:gg_rotations} and given by Eqs.~\eqref{eq:phase_shift_gg} and \eqref{eq:phase_shift_rotation},} depend linearly on $\hat{\mathbf{n}}$ and will also cancel out with this method. 
The remaining three terms in Eq.~\eqref{eq:corrections1}, on the other hand, do not cancel out, but can be neglected as long as $\Delta\bar{\mathbf{v}}_0$ and $\Delta\mathbf{g}$ are sufficiently small compared to $\bar{\mathbf{v}}_0$ and $\mathbf{g}$ respectively.

Note that although we have considered above a time-independent $\Delta\mathbf{g}$, the results can be straightforwardly generalized to the time-dependent case. In particular, the factors $\Delta\mathbf{g}\, T^2$ and $\Delta\mathbf{g}\, T^3$ in Eq.~\eqref{eq:corrections1} will then be replaced by double time integrals of $\Delta\mathbf{g} (\bar{t})$.
Such time dependence of $\Delta\mathbf{g}$ can be due to small time-dependent perturbations of the gravitational field. In that case the analog of the first term on the right-hand side of Eq.~\eqref{eq:corrections1} will still cancel out when employing the method described in the previous paragraph if the two atom interferometers with common mid-point trajectory but opposite $\hat{\mathbf{n}}$ are operated simultaneously. Furthermore, the simultaneous operation of the reversed interferometer will also enable the cancelation of phase-shift contributions due to gravity gradients and rotations that are sensitive to the initial conditions, even if there is initial-position and -velocity jitter from shot to shot.

\begin{figure}[t]
\begin{center}
\includegraphics[width=4.0cm]{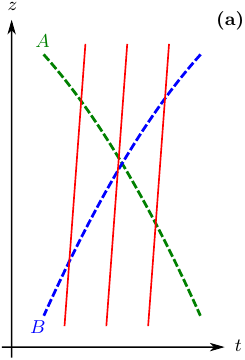}
\hspace{2.0ex}
\includegraphics[width=4.0cm]{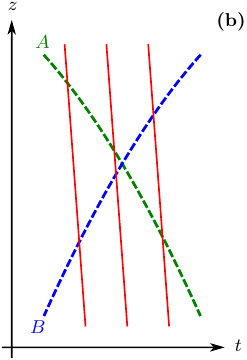}
\end{center}
\caption{Spacetime diagram in the laboratory frame that depicts the ``gradiometric'' configuration involving a pair of simultaneously operated atom interferometers with different initial velocities. The two atom clouds are independently launched from the top ($A$) and bottom ($B$) sources and interrogated by three common laser pulses consisting each of two slightly different frequencies so that both interferometers can be resonantly addressed. Only the central trajectories of the two interferometers (dashed lines) and central wave fronts of the laser pulses (continuous red lines) are shown. The pair of atom interferometers are interrogated by upward propagating pulses (a) whereas a pair of reversed interferometers are alternatively interrogated by downward propagating ones (b).}
\label{fig:gradiometric_conf}
\end{figure}

A time-dependent $\Delta\mathbf{g}$ can also account 
for phase fluctuations of the laser wave fronts due to laser phase noise or vibrations of the retro-reflection mirror, which both lead to a time-dependent $\mathbf{g}'$. However, these effects will in general be different for the two interferometers with opposite $\hat{\mathbf{n}}$ and their contributions will not cancel out when adding their phase shifts.
In order to address this point, one can use 
the \emph{gradiometric configuration} depicted in Fig.~\ref{fig:gradiometric_conf}, which involves two spatially separated interferometers ($A$ and $B$) with different initial velocities and interrogated by a common (possibly retro-reflected) laser beam. Indeed, for the differential phase shift $\delta\phi_A - \delta\phi_B$ the effects of laser phase noise and mirror vibrations will be common-mode rejected%
\footnote{\highlight{Due to the different velocities of the two interferometers, they are resonantly addressed by slightly different laser frequencies. Hence, there is only a partial common-mode rejection, but it still leads to a suppression factor $\left( \bar{\mathbf{v}}_0^A - \bar{\mathbf{v}}_0^B \right) / c \sim 10^{-7}$.}}.
Moreover, by considering a reversed pair of interferometers with the same mid-point trajectories but opposite $\hat{\mathbf{n}}$, as shown in Fig.~\ref{fig:gradiometric_conf}b, the remaining unwanted corrections linear in $\hat{\mathbf{n}}$ will cancel out when adding the differential measurements for the two signs of $\hat{\mathbf{n}}$, analogously to what happened for single interferometers. The final result, after taking the semisum of the differential phase shifts for opposite signs of $\hat{\mathbf{n}}$, is given by
\begin{equation}
\delta\phi_A - \delta\phi_B
= - 2\, (\Delta E / \hbar) \left( \bar{\mathbf{v}}_0^A - \bar{\mathbf{v}}_0^B \right) \cdot \mathbf{g}\, T^2 / c^2
\label{eq:gradiometric} ,
\end{equation}
where terms of order $1/c^2$ proportional to $\Delta\bar{\mathbf{v}}_0$ or $\Delta\mathbf{g}$ have not been included.

Using this gradiometric configuration also has favourable implications for the frequency chirp that should be applied to the laser pulses in order to compensate the Doppler factor. The required angular frequency $\omega_\text{chirp} (t) = (d \varphi / dt)_\text{chirp}$ can be obtained by inverting Eq.~\eqref{eq:chirp1}, substituting $\bar{\mathbf{v}}' (\bar{t}) = \bar{\mathbf{v}}'_0 + \mathbf{g}' \left( \bar{t} - \bar{t}_0 \right)$ and writing $(\bar{t} - \bar{t}_0)$ in terms of the emission time $t$ \comment{as explained in Appendix~\ref{sec:chirp_rate}.} The result contains terms of order $1/c^2$ which are proportional to $(\mathbf{g}')^2$ and depend quadratically on $(t - t_0)$. If a frequency $\omega_\text{chirp} (t)$ omitting such quadratic terms is employed instead, unwanted contributions of the same form arise in the phase shift $\delta\phi$.
Nevertheless, since those terms are independent of the initial velocity, they will cancel out in the differential phase shift, leaving the result in Eq.~\eqref{eq:gradiometric} unchanged. In this case it is therefore sufficient to use the following chirped frequency with purely linear dependence on time:
\begin{align}
\omega_\text{chirp} (t) &= \left[ 1 + \frac{ ( \hat{\mathbf{n}} \cdot \bar{\mathbf{v}}'_0 ) }{c}
+ \frac{ ( \hat{\mathbf{n}} \cdot \mathbf{g}' ) }{c} (t - t_0)
\right. \nonumber \\
& \left. \quad\ \ 
+ \frac{(\hat{\mathbf{n}} \cdot \bar{\mathbf{v}}'_0)^2}{c^2}
+ 3\, \frac{(\hat{\mathbf{n}} \cdot \bar{\mathbf{v}}'_0) \, (\hat{\mathbf{n}} \cdot \mathbf{g}')}{c^2} (t - t_0)
\right] \omega_0
\label{eq:chirp2b} .
\end{align}


\section{Experimental implementation}
\label{sec:implementation}

The measurement scheme proposed in the previous section can be naturally implemented in a long-baseline atom interferometry facility such as MAGIS-100, with a 100-m vertical baseline and three atomic sources (at the top, bottom and intermediate height) from which the atomic clouds can be independently
launched by means of accelerated optical lattices \cite{abe21}.
Specifically, we will consider the case in which two clouds of $^{87}\text{Sr}$ atoms are simultaneously launched from the top and bottom atomic sources. The bottom cloud is launched upwards and reaches a velocity $\bar{\mathbf{v}}_0 = (40\, \text{m/s})\, \hat{\mathbf{z}}$ at the time $t_0$ when the first beam-splitter pulse is applied, whereas the top cloud is launched downwards so that $\bar{\mathbf{v}}_0 = - (20\, \text{m/s})\, \hat{\mathbf{z}}$ at $t_0$.
For a total interferometer time $2 T = 2\, \text{s}$ and $\Delta E$ corresponding to the energy difference between the two clock states in Sr, we find from Eq.~\eqref{eq:gradiometric} a differential phase shift of \comment{$35\, \text{rad}$.} Hence, for $N=10^5$ detected atoms in each interferometer and a shot-noise limited measurement, the effect could be measured \comment{at the $10^{-5}$ level} in just a hundred shots.

The atom interferometers involve a sequence of three laser pulses driving the clock transition. The same kind of pulses can be employed for further preparing the initial state of the atoms after they have been launched. 
Indeed, by simultaneously applying a counter-propagating pair of velocity selection pulses
followed by a blow-away pulse, one can select from the original cloud two atomic clouds with much narrower momentum width and flying apart with a relative velocity corresponding to the recoil velocity $v_\text{rec}$. The two clouds can then be redirected with a second pair of counter-propagating pulses so that their central positions coincide when the first beam-splitter pulse of the interferometry sequence is applied. In this way, one can prepare the initial states for two 
interferometers with the same mid-point trajectory but opposite $\hat{\mathbf{n}}$ that can be simultaneously operated.
As additional preparation, the expansion in 
the transverse directions of the atomic clouds can be efficiently collimated with the matter-wave lensing technique already demonstrated in Ref.~\cite{kovachy15a}.

\highlight{
As explained in the previous section, by adding the phase shifts of the two interferometers with opposite $\hat{\mathbf{n}}$, the leading corrections associated with $\Delta \mathbf{g}$, gravity gradients and rotations cancel out.
Similarly, it can be shown that the systematic effects due to \emph{wave-front curvature} of the Gaussian laser beam are also suppressed provided that one focuses the beam waist at the retro-reflection mirror.
}

The laser frequency for one of the two atom interferometers with upward pointing $\hat{\mathbf{n}}$, as depicted in Fig.~\ref{fig:gradiometric_conf}a, can be generated with an acousto-optic modulator (AOM) driven by an RF signal that corresponds to the required frequency shift, including the linear chirp.
\comment{The additional frequency component needed for the second interferometer in the gradiometric configuration can be generated by mixing the RF signal that drives the AOM with a second signal that induces the additional frequency shift. Employing a highly stable RF source will guarantee the stability of the relative phase between the two frequency components in the laser beam and the suppression of any laser phase noise in the differential measurement.}
An analogous method can be used for the laser pulses with downward pointing $\hat{\mathbf{n}}$, displayed in Fig.~\ref{fig:gradiometric_conf}b, that \comment{interrogate} the pair of reversed interferometers
after being reflected by the retro-reflection mirror at the bottom, and guarantees that the effects of mirror vibrations are also suppressed in the differential phase shift.

The fermionic isotope $^{87}\text{Sr}$ will be employed because the clock transition is a forbidden transition for bosonic isotopes unless a strong magnetic field is applied. In contrast to the bosonic isotopes of alkali atoms, such as $^{87}\text{Rb}$, the total angular momentum of $^{87}\text{Sr}$ for the clock states comes entirely from the nuclear spin $I = 9/2$ (except for a very small hyperfine mixing of the excited clock state) \cite{boyd07}. As a result, the associated magnetic moments are \comment{about three orders of magnitude} smaller than for $^{87}\text{Rb}$ because of the large neutron-to-electron mass ratio. However, for a half-integer atomic spin the linear Zeeman effect is unavoidable. This can lead to non-negligible systematic effects due to \emph{inhomogeneities of the magnetic field}, but they can be effectively mitigated by \comment{alternating shots} with opposite signs of the magnetic quantum number $m_F$ (or even simultaneously realizing the two interferometers with opposite signs of $m_F$ in a single shot). Besides canceling out the linear Zeeman effect, the measurement outcomes for different values of $m_F$ can be exploited to infer the size of the magnetic-field inhomogeneities and confirm whether the contribution of the quadratic Zeeman effect is indeed negligible. \highlight{Further details about the systematic effects associated with magnetic fields can be found in Appendix~\ref{sec:magnetic_fields}.}

\highlight{
\emph{Temperature gradients} can also be a relevant source of systematic effects. Indeed,
the two clock states experience different AC Stark shifts in the presence of black-body radiation, which alters the accumulated relative phase between both states and constitutes an important systematic effect in (optical) atomic clocks. Due to the intermediate state inversion, the phase shift of the Mach-Zehnder interferometer will not be affected by a homogeneous (and time-independent) temperature background. However, temperature gradients will give rise to phase-shift contributions that resemble the gravitational effects of interest.
For a temperature of about $300\, \text{K}$ and a variation of $2\, \text{K}$
over $100\, \text{m}$, as considered for the facility proposed in Ref.~\cite{arduini23}, the relative size of the systematic contribution will be of the order of $1.3 \times 10^{-2}$.
By placing temperature sensors along the 100-m baseline to monitor such gradients, their effect can be  modeled and post-corrected, which may allow further reduction by at least one order of magnitude to the $10^{-3}$ level; see Appendix~\ref{sec:bb_radiation} for further details.
}

Finally, systematic effects associated with \emph{rotations} and \emph{gravity gradients} need to be considered as well. Interestingly, the leading phase-shift corrections, which are given by Eqs.~\eqref{eq:phase_shift_rotation} and \eqref{eq:phase_shift_gg}, are suppressed when considering the semisum of the two interferometers with opposite $\hat{\mathbf{n}}$
depicted in Fig.~\ref{fig:ff_frame}. In order to prepare such a pair of reversed interferometers, one can follow the approach outlined in the second paragraph of this section. 
However, any slight mismatch between the central positions of the \comment{atomic wave packets} when the first beam-splitter pulse is applied will lead to a residual contribution connected with gravity gradients, and similarly for rotations and gravity gradients if the difference of initial velocities does not exactly equal $\mathbf{v}_\text{rec}$. As an example, for the parameters specified in this section, measuring the relativistic time dilation with an accuracy at the $10^{-3}$ level would require matching the initial-position and -velocity differences with an accuracy better than $1.3\, \text{mm}$ and $1.3\, \text{mm/s}$ respectively, which can be straightforwardly achieved.
Moreover, if one chooses to have the two reversed interferometers in separate shots rather than simultaneously (e.g.~to avoid light shifts from the laser pulses of the other interferometer), shot-to-shot stability at that same level is also required.

Note also that the relative displacement between the interfering wave packets caused by rotations and gravity gradients can lead to loss of contrast if one simply measures the fraction of atoms detected at each exit port \cite{hogan08b,lan12,roura14,kleinert15,roura17a}. Nevertheless, such loss of contrast can be avoided by employing a phase-shear read-out technique \cite{sugarbaker13,muentinga13,roura14} provided that the relative displacement is much smaller than the spatial extent of the wave packet envelope.
For the parameters considered here Earth's rotation leads to \comment{a relative displacement of less than $0.5\, \mu\text{m}$} and the requirement is amply fulfilled -- the displacement due to gravity gradients is even smaller.
\highlight{Interestingly, by making use of the pivot-point method 
proposed in Ref.~\cite{glick24}, one may be able to extend to baselines of $100\, \text{m}$ the rotation compensation that has been successfully implemented in 10-m atomic fountains with a tip-tilt retro-reflecting mirror \cite{dickerson13}.
This can be employed to make sure that atomic clouds are located at the center of the laser beam when the pulses are applied despite the Coriolis acceleration.
Moreover, it would also suppress the main systematic effects due to rotations and relax the requirements on the co-location of the reversed interferometers 
because for $T = 1\, \text{s}$ the remaining effects due to gravity gradients are an order of magnitude smaller.
}

\section{Discussion}
\label{sec:discussion}

The central result of this paper, captured by Eqs.~\eqref{eq:phase_shift} and \eqref{eq:gradiometric}, is that light-pulse atom interferometers based on single-photon transitions can be used as freely falling clocks for the measurement of relativistic time-dilation effects. The higher precision that can be attained compared to standard atomic fountain clocks relies on the longer baseline available (up to a hundred meters rather than about a meter) and the higher energy difference between the two clock states, which corresponds to a transition frequency in the optical rather than the microwave regime.

A key aspect of the proposed measurement scheme is the suppression of the contribution from the Doppler \comment{shift}, which is seven orders of magnitude larger than the phase shift associated with the time-dilation effects. 
This can be achieved by combining two methods. Firstly, the frequency of the laser pulses is chirped  throughout the interferometer sequence at a suitable rate that approximately matches the gravitational acceleration. This approach, which is commonly employed in atomic-fountain interferometers \cite{peters01}, guarantees that the various laser pulses interrogating the freely falling atoms stay on resonance and it cancels out the main contribution of the \comment{Doppler factor.} Secondly, any remaining contribution due to an imperfect match of the chirp rate and the gravitational acceleration is further suppressed by \comment{(simultaneously)} operating a reversed 
interferometer with laser pulses propagating in the opposite direction and adding up the phase shifts of the two 
interferometers.

\highlight{
The second key aspect is the suppression of the undesirable effects of laser phase noise and mirror vibrations thanks to the ``gradiometric'' configuration depicted in Fig.~\ref{fig:gradiometric_conf}. In fact, the differential measurement can be regarded in that case as a direct comparison of the time-dilation effects for the two interferometers, $A$ and $B$, without the need for a highly accurate time reference in the laboratory frame. A stable local oscillatior for frequencies of several tens of MHz is still needed in order to drive the AOM that generates the additional laser frequency component so that both interferometers are resonantly addressed, 
but the requirements on its relative uncertainty are relaxed by more than seven orders of magnitude.
For example, an uncertainty of a few mHz is sufficient for a time-dilation measurement at the $10^{-4}$ level.}

The interferometry scheme considered here, which is conceptually closer to a clock following the mid-point trajectory, differs in many respects from quantum-clock interferometry \cite{sinha11,zych11,loriani19} experiments such as those proposed in Ref.~\cite{roura20a};  
see also Refs.~\cite{roura21,ufrecht20} for related variants.
In that case, atoms in a superposition of internal states and acting as clocks are prepared in a quantum superposition of two different heights. The difference in gravitational time dilation experienced along the two interferometer arms, which is reflected in the interference signal, is proportional to the height difference.
\comment{Hence, the measurement sensitivity is limited by the arm separation that can be achieved while keeping systematic effects under control.}
In contrast, the relevant length scale for the atom interferometer acting as a freely falling clock investigated here, which is given by $( \hat{\mathbf{n}} \cdot \bar{\mathbf{v}}_0 )\, T$, can be increased with the launch velocity and is mainly limited by the total baseline, namely 100 m for MAGIS-100.

In the previous sections we have focused on the general relativistic case, but the extension to
more general
frameworks that can consistently parametrize violations of the equivalence principle \cite{will14} is
discussed in \comment{Appendix~\ref{app:violations}.} The main conclusion \comment{there} is that the atom interferometric scheme can test the universality of gravitational redshift (UGR) in the same way that a freely falling clock following the mid-point trajectory would.
In any case, it is important to emphasize that we are dealing here with the measurement of a non-vanishing \comment{general} relativistic effect rather than a null test such as that considered in Ref.~\cite{di_pumpo23}.
Indeed, the experiments proposed in that reference could only search for differences in the clock rate between different isotopes of Sr or Yb. Furthermore, since they involve the comparison of 
different bosonic isotopes or a fermionic and bosonic pair,
the possibility of using interferometers based on single-photon transitions is excluded in practice because the clock transition is forbidden for bosonic isotopes unless a strong magnetic field is applied, which is not a viable option for precision measurements and long baselines.

\highlight{In summary, the atom interferometry scheme proposed here should enable the unprecedented measurement of gravitational time dilation in a local experiment with freely falling atoms by exploiting a long-baseline atom interferometric facility such as MAGIS-100 and with virtually no additional requirements. The proposal is particularly timely because MAGIS-100 \comment{is expected to start operating in just a few years}
and several other similar facilities, such as AION and ZAIGA, should follow after that. 
\comment{Furthermore, preliminary measurements with limited sensitivity will also be possible in the smaller-scale prototypes involving 10-m atomic fountains \cite{hogan-lab} that will become available very soon.}}


\acknowledgments{This work is supported by the Q-GRAV Project within the Space Research and Technology Program of the German Aerospace Center (DLR).}


\vspace{5.0ex}

\appendix

\appsection{Wave-packet propagation}
\label{app:propagation}

In order to obtain the evolution of the atomic wave packets, we will employ the relativistic description of matter-wave propagation in curved spacetime introduced in Ref.~\cite{roura20a}. By considering a suitable reference frame comoving with the matter-wave packet, its propagation can be conveniently described in terms of its \emph{central trajectory} and a centered wave packet
$\big| \psi_\text{c}^{(n)} (\tau_\mathrm{c}) \big\rangle$:
\begin{equation}
\big| \psi^{(n)} (\tau_\text{c}) \big\rangle = e^{i\hspace{0.2ex} \mathcal{S}_n / \hbar}\,
\big| \psi_\text{c}^{(n)} (\tau_\mathrm{c}) \big\rangle
\label{eq:state_evol1} ,
\end{equation}
where the index $n = 1,2$ 
labels the internal state and $\mathcal{S}_n$ is the \emph{propagation phase}. For freely falling atoms it is given by the rest-mass energy times the proper time along the central trajectory:
\begin{equation}
\mathcal{S}_n = - m_n c^2 \int^{\tau}_{\tau_0} d\tau'
\label{eq:prop_phase1} .
\end{equation}
As long as its size is much smaller than the spacetime curvature radius and its velocity spread is much smaller than the speed of light, the evolution of the \emph{centered wave packet} in the comoving frame is governed by a Schr\"odinger equation with a non-relativistic Hamiltonian $\hat{H}_\text{c}^{(n)} (\tau_\text{c})$ that includes the effects of spacetime curvature on the expansion dynamics of the centered wave packet:
\begin{equation}
i \hbar \frac{d}{d \tau_\text{c}} \big| \psi_\text{c}^{(n)} (\tau_\mathrm{c}) \big\rangle
=  \hat{H}_\text{c}^{(n)} (\tau_\text{c}) \, \big| \psi_\text{c}^{(n)} (\tau_\mathrm{c}) \big\rangle
\label{eq:Schroedinger} .
\end{equation}

Further details can be found in Ref.~\cite{roura20a}, where a relativistic description of atom interferometry in curved spacetime applicable to a wide range of situations, including also the effects of any external forces and guiding potentials, has been developed.
Throughout the present paper we will often make use of the fact that for non-relativistic velocities and weak gravitational fields the propagation phase $\mathcal{S}_n$ reduces to
\begin{equation}
\mathcal{S}_n \approx \int^{t}_{t_0} dt' \left( - m_n\, c^2 + \frac{1}{2} m_n \dot{\mathbf{X}}^2
- m_n\,  U (t',\mathbf{X}) \right)
\label{eq:prop_phase2} ,
\end{equation}
which coincides with the classical non-relativistic action plus a rest-mass energy contribution.


\subsection*{Quantum-clock evolution}

The results for matter-wave propagation summarized above can also be employed for describing the evolution of a quantum clock involving a superposition of two different internal states.
The associated Hilbert space is the tensor product of the internal space and the Hilbert space for the center-of-mass degree of freedom, so that the quantum state of the clock is given by
\begin{equation}
\big| \Psi \big\rangle = \big| \psi^{(1)} \big\rangle \otimes | \mathrm{g} \rangle
+ \big| \psi^{(2)} \big\rangle \otimes | \mathrm{e} \rangle
\label{eq:state1} .
\end{equation}

If the central trajectories for the two internal states are the same, the accumulated relative phase between them can be directly obtained from the difference of propagation phases for the two states:
\begin{equation}
\delta\phi = (\mathcal{S}_2 - \mathcal{S}_1) / \hbar = - (\Delta E / \hbar)\, \Delta\tau
\label{eq:clock_phase1} ,
\end{equation}
which is proportional to the proper time along the world line of the central trajectory
and where \comment{we have taken into account that the rest-mass difference $\Delta m$ between atoms in the two internal states is directly related to the energy difference between the two states: $\Delta m = \Delta E / c^2$.}
Therefore, if we neglect any differences in the evolution of the centered wave packets for the two states, whose justification is discussed in the last paragraph of this appendix,
the state of the quantum clock is given by  
\begin{equation}
\big| \Phi (\tau) \big\rangle \propto \frac{1}{\sqrt{2}}
\left( | \mathrm{g} \rangle + e^{- i \Delta E \, \Delta\tau / \hbar} | \mathrm{e} \rangle \right)
\label{eq:clock_state} ,
\end{equation}
clearly showing that the relative phase between the two internal states is \comment{directly connected to} the elapsed proper time $\Delta\tau$.

For non-relativistic velocities and weak gravitational fields, the proper time $\Delta\tau$ is well approximated by Eq.~\eqref{eq:proper_time}. One can then consider the case of a uniform gravitational field by taking $U (t',\mathbf{x}) = U_0 - \mathbf{g} \cdot (\mathbf{x} - \mathbf{x}_0)$ and substituting into Eq.~\eqref{eq:proper_time} 
the freely falling central trajectory
\begin{equation}
\mathbf{X}(t') = \mathbf{X}_0 + \mathbf{v}_0 (t' - t_0) + \mathbf{g}\, (t' - t_0)^2 / 2
\label{eq:ff_trajectory} .
\end{equation}
As a particular example, one can calculate $\Delta \tau$
for the freely falling clock shown in Fig.~\ref{fig:clock_comparison} by integrating from $t_0$ to $t_0 + T$ in  Eq.~\eqref{eq:proper_time} and taking $\mathbf{v}_0 = - \mathbf{g}\, T / 2$.
After substitution into Eq.~\eqref{eq:clock_phase1}, the result obtained for the accumulated relative phase is
\begin{equation}
\delta\phi = - (\Delta E / \hbar) \left( \left( 1 + U_0 / c^2 \right) T + \frac{1}{24} \frac{g^2 T^3}{c^2} \right)
\label{eq:ff-clock1_phase} .
\end{equation}

One can proceed analogously for a freely falling clock whose internal state is inverted at an intermediated time $t_0 + T$ 
and finally read out at a time $t_0 + 2 T$, as depicted in Fig.~\ref{fig:ideal_clock}. In this case, one needs to split the time integral into two parts and include a change of sign for the second integral, from $t_0 + T$ to $t_0 + 2 T$, in order to account for the 
inversion of the internal state. The resulting phase difference is given by
\begin{equation}
\delta\phi 
= - 2 \,(\Delta E / \hbar) \left( \mathbf{v}_0 \cdot \mathbf{g}\, T^2 + g^2 T^3  \right) / c^2
\label{eq:ff-clock2_phase} ,
\end{equation}
which corresponds to Eq.~\eqref{eq:clock_phase} in the main text.

Finally, note that the slight difference in the free expansion of the centered wave packets for the two internal states, which has been neglected above, amounts to small corrections of order $(\sigma_p / p_\text{rec})^2 (\omega_\text{rec} T) (\Delta m / m)$ to Eq.~\eqref{eq:ff-clock1_phase}, where $\sigma_p$ is the momentum width of the wave packet and the recoil frequency $\omega_\text{rec} = (p_\text{rec}^2 / 2m) / \hbar$ has been introduced for convenience and later comparison.
For $\sigma_p \lesssim 0.1\, p_\text{rec}$ with $p_\text{rec}$ corresponding to the momentum of a photon in the optical regime and $T \sim 1 \, \text{s}$,
the size of these corrections is below $10^{-8}\, \text{rad}$.
Moreover, when considering a state inversion at the intermediate time, which leads to the result of Eq.~\eqref{eq:ff-clock2_phase}, this small correction cancels out
and only higher-order contributions, which are even smaller, remain.

\appsection{Freely falling frame}
\label{app:ff_frame}

In order to calculate the phase shift for an atom interferometer, one needs to calculate the light-ray propagation in the curved spacetime under consideration as well as the time-like geodesics corresponding to the central trajectories of the atomic wave packets and the proper time along these. Interestingly, these tasks can be significantly simplified by considering a suitable freely falling frame. More specifically, we will consider the Fermi-Walker frame associated with the mid-point trajectory between the two interferometer arms, 
and the corresponding Fermi coordinates. In this coordinate system the world line for the mid-point trajectory reduces to $\bar{X}^\mu(t_\text{FW}) = \big(c\, t_\text{FW}\, , \mathbf{0} \big)$ and the metric is given by the following line element:
\begin{align*}
ds^2 &= g_{\mu\nu} dx^\mu dx^\nu
\nonumber \\
&= g_{00}\, c^2 dt_\text{FW}^2
\,+\, 2\, g_{0i}\, c\, dt_\text{FW}\, dx^i
\,+\, g_{ij}\, dx^i dx^j ,
\end{align*}
with components
\begin{align}
g_{00} &= -1 
- R_{0i0j} (t_\text{FW},\mathbf{0})\, x^i x^j 
+ O\big( |\mathbf{x}|^3 \big) , \label{eq:FW_metric2a} \\
g_{0i} &= - \frac{2}{3} R_{0jik} (t_\text{FW},\mathbf{0})\, x^j x^k
+\, O\big( |\mathbf{x}|^3 \big) ,  \label{eq:FW_metric2b} \\
g_{ij} & = \delta_{ij} - \frac{1}{3} R_{ikjl} (t_\text{FW},\mathbf{0})\, x^k x^l
+\, O\big( |\mathbf{x}|^3 \big) , \label{eq:FW_metric2c}
\end{align}
where \comment{the time coordinate $t_\text{FW}$ coincides with the proper time $\bar{\tau}$ along the world line and $R_{abcd}$ are the components of the Riemann tensor, which characterizes the spacetime curvature.}
The metric defined by Eqs.~\eqref{eq:FW_metric2a}--\eqref{eq:FW_metric2c} reduces to the Minkowski metric along the world line and gets corrections due to the spacetime curvature as one moves away from it. A more detailed discussion can be found in Ref.~\cite{roura20a} and references therein.

\subsection{Locally uniform gravitational field}
\label{sec:uniform_field}

We will first focus on the case where gravity gradients (or, equivalently, spacetime curvature) can be neglected over length scales comparable to the separation between the interferometer arms. The metric in Eqs.~\eqref{eq:FW_metric2a}--\eqref{eq:FW_metric2c} reduces then to the Minkowski metric and the central trajectories for the freely falling atomic wave packets correspond to straight world lines. Light rays will also follow straight lines, as depicted in the spacetime diagrams of Figs.~\ref{fig:ff_frame}a and \ref{fig:ff_frame}b. In particular, light rays along the \comment{$z$ direction} will all have the same slope but will be shifted due to Doppler and gravitational redshift effects.

In order to determine these shifts, one needs to calculate the intersection of the light rays with the mid-point world line $\bar{X}^\mu(t) = \big( c\, t, \bar{\mathbf{X}}(t) \big)$ 
from the point of view of the laboratory frame.
When doing so, we will consider a post-Newtonian expansion in a \comment{static} spacetime, with time coordinate $t$ associated with the spacetime's time-translation invariance.
In that case the time separation $dt$ between two infinitesimally close light rays propagating along the same direction $\hat{\mathbf{n}}$ will remain constant.
However, due to the motion of the freely falling mid-point trajectory, the time difference $d\bar{t}$ between its intersection with the two light rays will satisfy $d\bar{t} = dt + (\hat{\mathbf{n}} \cdot \bar{\mathbf{v}} / c)\, d\bar{t}$, where $\bar{\mathbf{v}} = d \bar{\mathbf{X}} / d \bar{t}$
\comment{and terms of order $1/c^3$ have been neglected.}
Therefore, the times for different light rays at a fixed position in the laboratory frame and the intersection times with the mid-point trajectory are connected by the following differential relation:
\begin{equation}
\quad \ 
\frac{d \bar{t}}{d t} = \frac{1}{1 - \hat{\mathbf{n}} \cdot \bar{\mathbf{v}} / c}
\label{eq:Doppler2} ,
\end{equation}
which corresponds to the ``classical'' Doppler effect
\highlight{arising from the retardation effects due to the finite speed of light and the motion of the atoms with respect to the laboratory frame.}

The times considered in the previous paragraph are all connected to the time coordinate associated with time-translation invariance in the laboratory frame. On the other hand, for non-relativistic velocities and weak gravitational fields the proper time along the mid-point world line is given by
\begin{equation}
\frac{d \bar{\tau}}{d \bar{t}} \,=\, 1 - \frac{1}{2\, c^2}\! \left( \frac{d \bar{\mathbf{X}}}{d \bar{t}} \right)^2
+\, \frac{1}{c^2}\, U \big( \bar{t},\bar{\mathbf{X}} \big)
+ O \big( 1/c^4 \big)
\label{eq:dilation} ,
\end{equation}
which includes both special relativistic and gravitational time-dilation effects.
The time separation 
between the two light rays in the Fermi-Walker frame corresponds to the proper time calculated along the mid-point world line and elapsed between the intersection points with the two light rays. It is therefore related to the difference of emission times in the laboratory frame through the following expression:
\begin{equation}
\quad \ 
\frac{d \bar{\tau}}{d t} =  \frac{d \bar{t}}{d t}\, \frac{d \bar{\tau}}{d \bar{t}}
\label{eq:} ,
\end{equation}
with the two factors on the right-hand side given by Eqs.~\eqref{eq:Doppler2} and \eqref{eq:dilation} respectively.

\subsection{Non-uniform gravitational field}
\label{sec:non-uniform_field}

Non-uniform gravitational fields are associated with a non-vanishing spacetime curvature, characterized by the Riemann tensor $R_{abcd}$. For objects moving at non-relativistic speeds the particularly relevant components of the Riemann tensor are the temporal ones, which are directly connected to the gravity gradient tensor $\Gamma_{ij}$ through the relation $\Gamma_{ij}  = - c^2 R_{0i0j}$. 
To lowest order in a post-Newtonian expansion
the gravity gradient tensor is, in turn, given by the Hessian of the Newtonian potential: $\Gamma_{ij} = -\partial^2 U / \partial x^i \partial x^j$.

Deviations from a uniform field have an impact on the result for the mid-point trajectory calculated in the laboratory frame. For a time-independent gravity gradient there is an exact analytical solution in the Newtonian regime \cite{roura14}, but it is often convenient to consider a perturbative expansion in powers of $\left( \Gamma (\Delta\bar{t})^2 \right)$:
\begin{align}
\bar{\mathbf{X}}(\bar{t}) &= \bar{\mathbf{X}}_0 +  \bar{\mathbf{v}}_0\, \Delta\bar{t}
+ \frac{1}{2}\, \mathbf{g}\, (\Delta\bar{t})^2
\nonumber \\
& \quad + \frac{1}{2} \left( \Gamma\, (\Delta\bar{t})^2 \right)
\left( \bar{\mathbf{X}}_0 + \frac{1}{3} \bar{\mathbf{v}}_0\, \Delta\bar{t}
+ \frac{1}{12}\, \mathbf{g}\, (\Delta\bar{t})^2 \right)
\label{eq:trajectory_gg} ,
\end{align}
where $\Delta\bar{t} = (\bar{t} - \bar{t}_0)$ and we have neglected terms involving higher powers of $\left( \Gamma (\Delta\bar{t})^2 \right)$. The well-known result for the uniform case is clearly recovered when taking $\Gamma = 0$.
\comment{Moreover, for a time-dependent gravity gradient tensor $\Gamma (\bar{t})$ there is a straightforward generalization of Eq.~\eqref{eq:trajectory_gg} where the factor $(\Delta\bar{t})^2$ is replaced by a double time integral \cite{roura14,kleinert15}.}

In addition, the tidal forces associated with gravity gradients, and with the curvature term in Eq.~\eqref{eq:FW_metric2a}, lead to deviations for the central trajectories of the atomic wave packets in the freely falling frame, which are no longer given by simple straight lines in the spacetime diagram.
In fact,
since the component $\Gamma_{zz}$ is positive for Earth's gravitational field, the associated tidal forces tend to open up the spacetime trajectories for motions along the vertical direction.
The corresponding trajectories 
can be directly obtained by taking $\mathbf{g} = \mathbf{0}$ in Eq.~\eqref{eq:trajectory_gg}, replacing the time coordinate $\bar{t}$ with $t_\text{FW}$ and considering the corresponding initial conditions for each segment of the central trajectories.
Calculating in this way the central trajectories for the two interferometer arms, one finds the following
result to leading order in $\Gamma$ and $1/c$ for the 
relative displacement between the two interfering wave packets at each exit port:
\highlight{
\begin{equation}
\delta \boldsymbol{\mathbf{X}} = \left(\Gamma\, T^2 \right) \mathbf{v}_\text{rec} \, T \, ,
\qquad
\delta \boldsymbol{\mathbf{P}} = (\Gamma\, T^2) \, m\, \mathbf{v}_\text{rec}
\label{eq:displacements_gg} \, ,
\end{equation}
where $\mathbf{v}_\text{rec}$ 
is the recoil velocity associated with the single-photon momentum transfer, given by
\begin{equation}
\mathbf{v}_\text{rec} = \frac{\Delta E}{m\,c}\, \hat{\mathbf{n}}
\,\, \Big( 1 + O \big( \Delta E / m c^2 \big) \Big)
\label{eq:recoil_vel} ,
\end{equation}
and further discussed in Appendix~\ref{sec:recoil_vel}.
Such relative displacements result} in sensitivity of the interference signal to 
the initial position and velocity, as can be seen in the phase-shift corrections due to gravity gradients obtained in Appendix~\ref{app:gg_rotations}.

Finally, although spacetime curvature also modifies the spacetime trajectories of light rays, the effect is much smaller than for non-relativistic particles due to light's far shorter time of flight between the two interferometer arms. Indeed, compared to the effects on the trajectories of the atomic wave packets, which are of order $(\Gamma\, T^2)$, the effects on light rays are further suppressed by a factor $(v_\text{rec} / c)^2 \sim 10^{-22}$ and are completely negligible in this context.

\appsection{Phase-shift calculation}
\label{app:phase-shift}

We are now ready to calculate the phase shift for a Mach-Zehnder interferometer making use of the elements introduced in the previous appendices. As explained in Appendix~\ref{app:ff_frame}, it is convenient to consider a suitable freely falling frame, namely the Fermi-Walker frame associated with the mid-point trajectory between the two interferometer arms. Moreover, we will focus here on the case of a uniform gravitational field, whereas the effects of gravity gradients will be discussed in Appendix~\ref{app:gg_rotations}.

To obtain the phase shift $\delta\phi$, we need to compute the propagation phase
along each interferometer arm and take the difference between the two arms. It is clear from Fig.~\ref{fig:ff_frame} that in the absence of any Doppler and time-dilation 
effects the total propagation phase along the two arms is the same.
\comment{This is because in that case the central trajectories for the two arms are point symmetric with respect to the intersection of the mid-point world line and the central wave front of the second laser pulse; see Appendix~\ref{sec:recoil_vel} for the detailed definition of the mid-point world line.}
On the other hand, as explained in Appendix~\ref{sec:uniform_field}, Doppler and time-dilation effects will both shift light rays in the spacetime diagram while keeping their slope. As a result, the proper time spent by the atoms in the excited state in the first and second half of the interferometer (i.e.\ before and after the second laser pulse) will be different. \comment{Interestingly, this proper-time difference can be 
obtained (up to subleading corrections) from the proper times calculated along the mid-point world line for the segments delimited by its intersections with the shifted light rays,
as shown next and illustrated by Fig.~\ref{fig:ff_frame}.} 

Note that since the phase $\varphi$ of any given laser wave front (corresponding to a null hypersurface in spacetime) is a frame-independent quantity, it is convenient to express \comment{the intermediate results} in terms of this variable. When doing so, we will assume \comment{for the moment} that the laser phase and the timing of the laser pulses (their envelope) are tied together;
deviations from this assumption will be discussed in Appendix~\ref{sec:pulse_timings}.
The interferometer phase shift $\delta\phi$, which is proportional to $\Delta m$ and the difference between the times spent in the excited state along the two arms, can then be written as
\begin{equation}
\delta\phi = - \frac{\Delta E}{\hbar}
\left[
\int_0^{\omega_0 T} \! \left( \frac{d \bar{\tau}}{d \varphi} \right) \,  d\varphi
\, - \int_{\omega_0 T}^{2 \omega_0 T}  \! \left( \frac{d \bar{\tau}}{d \varphi} \right) \,  d\varphi
\right]
\label{eq:phase_shift0} ,
\end{equation}
where $\bar{\tau}$ is the proper time along the mid-point trajectory and we have taken into account that
$\Delta m = \Delta E / c^2$.
\comment{Calculating the proper times along the central trajectories of the two arms rather than the mid-point trajectory would give rise to extra contributions with additional powers of $v_\text{rec} / c$ and $v_\text{rec}^2 / c^2$ multiplying the terms that contribute to Eq.~\eqref{eq:phase_shift0} and are already of order $1/c^2$.
A detailed analysis showing that any differences between the evaluation of the propagation phases along the arm trajectories and along the mid-point world line can be safely neglected is provided in Appendix~\ref{app:arm_trajectories}.}

In order to compute the right-hand side of Eq.~\eqref{eq:phase_shift0}, one needs to consider the relation between the laser phase $\varphi$ and the times \comment{$t$ or $\bar{t}$} in the laboratory frame. This point will be analyzed 
in the next two subsections.

\subsection{Perfect cancelation of the Doppler factor}
\label{sec:Doppler_cancelation}

For a static light source in the laboratory frame emitting with constant angular frequency $\omega_0$, time and phase satisfy the simple relation $d \varphi / dt = \omega_0$. However, due to the motion of the freely falling atoms, there will be a Doppler shift of the time separations between laser wave fronts intersecting the mid-point trajectory, as determined by Eq.~\eqref{eq:Doppler2}. By chirping the frequency according to Eq.~\eqref{eq:chirp1} and choosing the appropriate chirp rate, the Doppler factor on the right-hand side of Eq.~\eqref{eq:proper-time_phase} can be compensated and one is left with $(d \bar{t} / d \varphi) = 1 / \omega_0$, which corresponds to equal time separations between laser wave fronts at the mid-point trajectory.

Assuming such a perfect cancellation of the Doppler factor, one can trivially replace the phase $\varphi$ with the laboratory time $\bar{t}$ along the mid-point trajectory, and Eq.~\eqref{eq:phase_shift0} becomes
\begin{equation}
\delta\phi = - \frac{\Delta E}{\hbar}
\left[
\int_0^{T} \! \left( \frac{d \bar{\tau}}{d \bar{t}} \right) \,  d \bar{t}
\, - \int_T^{2 T} \! \left( \frac{d \bar{\tau}}{d \bar{t}} \right) \,  d \bar{t}
\right]
\label{eq:phase_shift1} .
\end{equation}
where the integrand corresponds to the time-dilation factor of Eq.~\eqref{eq:dilation} and 
\comment{we have chosen the origin of $\bar{t}$ so that $\bar{t}_0 = 0$.}
The result coincides with that for an ideal clock following the mid-point trajectory and undergoing a recoilless inversion of the internal state at the intermediate laboratory time.
Indeed, substituting Eq.~\eqref{eq:dilation} into Eq.~\eqref{eq:phase_shift1} and taking into account that
\begin{equation}
\bar{\mathbf{X}}(\bar{t}) =  \bar{\mathbf{X}}_0 +  \bar{\mathbf{v}}_0\, (\bar{t} - \bar{t}_0)
+ \mathbf{g}\, (\bar{t} - \bar{t}_0)^2 / 2
+ O (1/c^2 )
\label{eq:trajectory3} ,
\end{equation}
\comment{where higher-order terms in the post-Newtonian expansion have been omitted,}
we obtain the following result for the phase shift of the Mach-Zehnder interferometer:
\begin{equation}
\delta\phi 
= - 2 \,(\Delta E / \hbar) \left( \bar{\mathbf{v}}_0 \cdot \mathbf{g}\, T^2 + g^2 T^3  \right) / c^2
\label{eq:phase_shift2} ,
\end{equation}
which agrees with the result obtained in Appendix~\ref{app:propagation} for an ideal clock and corresponding to Eq.~\eqref{eq:ff-clock2_phase}.

\vspace{0.0ex}

\subsection{Incomplete cancelation of the Doppler factor}
\label{sec:incomplete_Doppler_cancelation}

In practice, the frequency chirp will not compensate the Doppler factor completely. More specifically, the parameters $\bar{\mathbf{v}}'_0$ and $\mathbf{g}'$ determining $\bar{\mathbf{v}}'$ in Eq.~\eqref{eq:chirp1} will not exactly coincide with $\bar{\mathbf{v}}_0$ and $\mathbf{g}$. Instead, there will be small differences $\Delta \bar{\mathbf{v}}_0 = \bar{\mathbf{v}}_0 - \bar{\mathbf{v}}'_0$ and $\Delta\mathbf{g} = \mathbf{g} - \mathbf{g}'$.
Therefore, when multiplying Eqs.~\eqref{eq:Doppler} and \eqref{eq:chirp1}, one is left with the following result:
\begin{widetext}

\begin{align}
\left( \frac{d \bar{t}}{d \varphi} \right)_\text{chirp}
= & \ \ \frac{1}{\omega_0} \left[ 1 + \frac{(\hat{\mathbf{n}} \cdot \Delta \bar{\mathbf{v}}_0)}{c}
+ \frac{(\hat{\mathbf{n}} \cdot \Delta\mathbf{g})}{c} (\bar{t} - \bar{t}_0)
+ \frac{(\hat{\mathbf{n}} \cdot \bar{\mathbf{v}}_0) \, (\hat{\mathbf{n}} \cdot \Delta\bar{\mathbf{v}}_0)}{c^2}
+ \frac{(\hat{\mathbf{n}} \cdot \Delta \bar{\mathbf{v}}_0) \,
(\hat{\mathbf{n}} \cdot \mathbf{g})}{c^2} (\bar{t} - \bar{t}_0) 
\right. \nonumber \\
& \qquad\qquad\qquad\qquad\qquad\qquad\qquad\qquad\ \ \left. +\, \frac{(\hat{\mathbf{n}} \cdot \bar{\mathbf{v}}_0) \, (\hat{\mathbf{n}} \cdot \Delta\mathbf{g})}{c^2} (\bar{t} - \bar{t}_0) 
+ \frac{(\hat{\mathbf{n}} \cdot \mathbf{g}) \,
(\hat{\mathbf{n}} \cdot \Delta\mathbf{g})}{c^2} (\bar{t} - \bar{t}_0)^2
\right]
\label{eq:corrections2} ,
\end{align}

\end{widetext}
where we have neglected terms of order $1/c^3$ or higher.

By integrating Eq.~\eqref{eq:corrections2}, one can directly obtain $\varphi$ in terms of $\bar{t}$. In principle, one would then need to invert this relation and express $\bar{t}$ in terms of $\varphi$, which can be done through a perturbative expansion, so as to calculate the right-hand side of Eq.~\eqref{eq:phase_shift0}. However, at the order that we are working this is not necessary if one proceeds as follows. First, one notes that the integrand in Eq.~\eqref{eq:phase_shift0} can be written as
\begin{equation}
\left( \frac{d \bar{\tau}}{d \varphi} \right)_\text{chirp} = \left( \frac{d \bar{\tau}}{d \bar{t}} \right)
\left( \frac{d \bar{t}}{d \varphi} \right)_\text{chirp}
\label{eq:_} ,
\end{equation}
where the two factors on the right-hand side are given by Eqs.~\eqref{eq:dilation} and \eqref{eq:corrections2} respectively. The product of the lowest-order contributions from both factors amounts to $1/\omega_0$ and for such a constant term the two integrals on the right-hand side of Eq.~\eqref{eq:phase_shift0} cancel out.
On the other hand, for all the other contributions it is actually enough to employ the lowest-order relation between $\bar{t}$ and $\varphi$, which is simply given by $\bar{t} - \bar{t}_0 = (\varphi - \varphi_0) / \omega_0$.
Indeed, since the higher-order terms in Eq.~\eqref{eq:dilation} are already of order $1/c^2$, it is sufficient \comment{in that case} to take $(d \bar{t} / d \varphi)_\text{chirp} \approx 1/\omega_0$ and consider the lowest-order relation between $\bar{t}$ and $\varphi$. The resulting phase-shift contribution reduces then to the right-hand side of Eq.~\eqref{eq:phase_shift1} and leads to the result in Eq.~\eqref{eq:phase_shift2}.

Similarly, since the higher-order terms in Eq.~\eqref{eq:corrections2} are of order $1/c$ or higher and proportional to $\Delta \bar{\mathbf{v}}_0$ or $\Delta\mathbf{g}$, it is enough to take $(d \bar{\tau} / d \bar{t}) \approx 1$ and consider again the lowest-order relation between $\bar{t}$ and $\varphi$.
\highlight{In doing so, one neglects terms quadratic in $\Delta \bar{\mathbf{v}}_0$ or $\Delta\mathbf{g}$.}
The resulting phase-shift contribution corresponds to the right-hand side of Eq.~\eqref{eq:phase_shift1}, but with $(d \bar{\tau} / d \bar{t})$ replaced by $(d \bar{t} / d \varphi)_\text{chirp}$:
\begin{align}
\delta\phi'_\text{corr} =& - \frac{\Delta E}{\hbar}
\left[
\int_0^{T} \! \left( \frac{d \bar{t}}{d \varphi} \right)_\text{chirp} \omega_0\, d \bar{t}
\right. \nonumber \\
& \left. \qquad\qquad
- \int_T^{2 T} \! \left( \frac{d \bar{t}}{d \varphi} \right)_\text{chirp} \omega_0\, d \bar{t}\,
\right]
\label{eq:corrections3a} .
\end{align}
Evaluating then the integral over $\bar{t}$ leads to the following phase-shift correction due to the incomplete cancellation of the Doppler factor:
\begin{align}
\delta\phi'_\text{corr} &= \frac{\Delta E}{\hbar} \, \left[ \frac{(\hat{\mathbf{n}} \cdot \Delta\mathbf{g})}{c}\, T^2
+ \frac{(\hat{\mathbf{n}} \cdot \Delta \bar{\mathbf{v}}_0) \, (\hat{\mathbf{n}} \cdot \mathbf{g})}{c^2}\, T^2
\right. \nonumber \\
& \qquad \left. +\, \frac{(\hat{\mathbf{n}} \cdot \bar{\mathbf{v}}_0) \, (\hat{\mathbf{n}} \cdot \Delta\mathbf{g})}
{c^2}\, T^2
+ 2\, \frac{(\hat{\mathbf{n}} \cdot \mathbf{g})\, (\hat{\mathbf{n}} \cdot \Delta\mathbf{g})}{c^2}\, T^3
\right]
\label{eq:corrections3b} .
\end{align}
A prime has been used to distinguish this result from a related result in Eq.~\eqref{eq:corrections4}, obtained below for a frequency chirp in terms of the laboratory time $t$.

\subsection{Chirp rate}
\label{sec:chirp_rate}

The frequency chirp required for compensating the Doppler factor,
which follows from inverting Eq.~\eqref{eq:chirp1} and is given by
\begin{equation}
\omega_\text{chirp} (t) = \big( 1 - \hat{\mathbf{n}} \cdot \bar{\mathbf{v}}' / c \big)^{-1} \omega_0
\label{eq:chirp2} ,
\end{equation}
is naturally expressed in terms of the time $\bar{t}$. In contrast, 
the frequency chirp applied to the static laser source is directly related to the laboratory time $t$. Therefore, we need to transform the time dependence of the chirp factor from the $\bar{t}$ to the $t$ variable.
The relation between these two time coordinates can be obtained by considering the mid-point trajectory given by Eq.~\eqref{eq:trajectory3} and substituting $\bar{\mathbf{v}} = d \bar{\mathbf{X}} / d \bar{t}$ into Eq.~\eqref{eq:Doppler2}. Integrating the resulting equation, one gets
\begin{equation}
(t - t_0) =  (\bar{t} - \bar{t}_0) \left( 1 - \frac{ ( \hat{\mathbf{n}} \cdot \bar{\mathbf{v}}_0 ) }{c}
- \frac{ ( \hat{\mathbf{n}} \cdot \mathbf{g} ) }{2 c} (\bar{t} - \bar{t}_0) \right)
\label{eq:time_transformation1} ,
\end{equation}
which can be inverted perturbatively and gives
\begin{align}
(\bar{t} - \bar{t}_0) &= (t - t_0) \left( 1 + \frac{ ( \hat{\mathbf{n}} \cdot \bar{\mathbf{v}}_0 ) }{c}
+ \frac{ ( \hat{\mathbf{n}} \cdot \mathbf{g} )\, (t - t_0) }{2 c}  \right)
\nonumber \\
&\quad + O \big( 1/c^2 \big)
\label{eq:time_transformation2} .
\end{align}
This result can then be substituted into Eq.~\eqref{eq:chirp2} 
so as to obtain the frequency chirp in terms of the time coordinate $t$.
When doing so, it is convenient to expand first the right-hand side of Eq.~\eqref{eq:chirp2} in powers of $1/c$:
\begin{widetext}

\begin{align}
\omega_\text{chirp} (t) &\approx \left[ 1 + \frac{\hat{\mathbf{n}} \cdot \bar{\mathbf{v}}'}{c}
+ \left( \frac{\hat{\mathbf{n}} \cdot \bar{\mathbf{v}}'}{c} \right)^2 \right] \omega_0
\approx \left[ 1 + \frac{ ( \hat{\mathbf{n}} \cdot \bar{\mathbf{v}}'_0 ) }{c}
+ \frac{(\hat{\mathbf{n}} \cdot \bar{\mathbf{v}}'_0)^2}{c}
+ \frac{ ( \hat{\mathbf{n}} \cdot \mathbf{g}' ) }{c} (t - t_0)
+ 2\, \frac{(\hat{\mathbf{n}} \cdot \bar{\mathbf{v}}'_0) \, (\hat{\mathbf{n}} \cdot \mathbf{g}')}{c^2} (t - t_0)
\right. \nonumber \\
& \left. \qquad\qquad\qquad\qquad\qquad\qquad\qquad
+ \frac{(\hat{\mathbf{n}} \cdot \mathbf{g}')^2}{c} (t - t_0)^2
+ \frac{(\hat{\mathbf{n}} \cdot \bar{\mathbf{v}}_0) \, (\hat{\mathbf{n}} \cdot \mathbf{g}')}{c^2} (t - t_0)
+ \highlight{\frac{1}{2}}
\frac{(\hat{\mathbf{n}} \cdot \bar{\mathbf{g}}) \, (\hat{\mathbf{n}} \cdot \mathbf{g}')}{c^2} (t - t_0)^2
\right] \omega_0 \,
\label{eq:chirp3} ,
\end{align}

\end{widetext}
where we have used $\bar{\mathbf{v}}' = \bar{\mathbf{v}}'_0 + \mathbf{g}'\, (\bar{t} - \bar{t}_0)$ as well as Eq.~\eqref{eq:time_transformation2} in the second equality, and we have neglected in all cases terms of order $1/c^3$ or higher.

Since one has direct control on the parameters \comment{$\bar{\mathbf{v}}'_0$ and $\mathbf{g}'$, which respectively characterize the initial frequency shift and the chirp rate,} it is more natural to consider $\bar{\mathbf{v}}'_0 = \bar{\mathbf{v}}_0 - \Delta \bar{\mathbf{v}}_0$ instead of $\bar{\mathbf{v}}_0$ on the right-hand side of Eq.~\eqref{eq:chirp3}, and $\mathbf{g}' = \mathbf{g} - \Delta \mathbf{g}$ instead of $\mathbf{g}$. By reversing the derivation of Eq.~\eqref{eq:chirp3}, one can see that these choices give rise to additional contributions to the right-hand side of Eq.~\eqref{eq:corrections2} that amount to a factor 2 and 3/2 for the terms proportional to $(\hat{\mathbf{n}} \cdot \Delta \bar{\mathbf{v}}_0) \, (\hat{\mathbf{n}} \cdot \mathbf{g})$ and $(\hat{\mathbf{n}} \cdot \mathbf{g})\, (\hat{\mathbf{n}} \cdot \Delta\mathbf{g})$, respectively,
\comment{as well as terms of higher order in $\Delta \bar{\mathbf{v}}_0$ and $\Delta \mathbf{g}$.}
In turn, such factors of 2 and 3/2 lead to analogous changes in the phase-shift correction in Eq.~\eqref{eq:corrections3b}, which becomes
\begin{align}
\delta\phi_\text{corr} &= \frac{\Delta E}{\hbar} \, \left[ \frac{(\hat{\mathbf{n}} \cdot \Delta\mathbf{g})}{c}\, T^2
+ 2\,
\frac{(\hat{\mathbf{n}} \cdot \Delta \bar{\mathbf{v}}_0) \, (\hat{\mathbf{n}} \cdot \mathbf{g})}{c^2}\, T^2
\right. \nonumber \\
& \qquad \left. +\, \frac{(\hat{\mathbf{n}} \cdot \bar{\mathbf{v}}_0) \, (\hat{\mathbf{n}} \cdot \Delta\mathbf{g})}
{c^2}\, T^2
+ 3\,
\frac{(\hat{\mathbf{n}} \cdot \mathbf{g})\, (\hat{\mathbf{n}} \cdot \Delta\mathbf{g})}{c^2}\, T^3
\right]
\label{eq:corrections4} .
\end{align}

The right-hand side of Eq.~\eqref{eq:chirp3} includes terms linear in time, which correspond to a constant chirp rate, but also terms that depend quadratically on time. Considering a chirped frequency where such quadratic terms are exlcuded, leads to an extra contribution proportional to $(\hat{\mathbf{n}} \cdot \mathbf{g}')^2\, T^3 / c^2$ to the phase-shift result in Eq.~\eqref{eq:phase_shift2}. Nevertheless, this extra term will cancel out when considering the differential phase shift between two identical interferometers with different initial velocities, which corresponds to Eq.~\eqref{eq:phase_shift2}.
In that case it is sufficient to employ the following frequency chirp:
\begin{align}
\omega_\text{chirp} (t) &= \left[ 1 + \frac{ ( \hat{\mathbf{n}} \cdot \bar{\mathbf{v}}'_0 ) }{c}
+ \frac{ ( \hat{\mathbf{n}} \cdot \mathbf{g}' ) }{c} (t - t_0)
\right. \nonumber \\
& \left. \quad\ \ 
+ \frac{(\hat{\mathbf{n}} \cdot \bar{\mathbf{v}}'_0)^2}{c^2}
+ 3\, \frac{(\hat{\mathbf{n}} \cdot \bar{\mathbf{v}}'_0) \, (\hat{\mathbf{n}} \cdot \mathbf{g}')}{c^2} (t - t_0)
\right] \omega_0
\label{eq:chirp4} ,
\end{align}
where the sign of the terms linear in $\hat{\mathbf{n}}$ changes for the \comment{reversed interferometer.}

\subsection{Phase noise and finite pulse duration}
\label{sec:phase_noise}

The results obtained in the previous two subsections for a time-independent $\Delta\mathbf{g}$ can be easily generalized to the time-dependent case. The generalization implies the following substitution in the three different terms on the right-hand side of Eq.~\eqref{eq:corrections2} where it appears:
\begin{equation}
\frac{ ( \hat{\mathbf{n}} \cdot \Delta\mathbf{g} ) }{c}\, (\bar{t} - \bar{t}_0)
\,\to\, \int^{\bar{t}}_{\bar{t}_0} d\bar{t}' \,
\frac{ \big( \hat{\mathbf{n}} \cdot \Delta\mathbf{g} \left( \bar{t}' \right) \big) }{c}
\label{eq:time_dependent} .
\end{equation}
Similarly, a factor of $T^2$ will be replaced by a double time integral in each of the three terms involving $\Delta\mathbf{g}$ in Eqs.~\eqref{eq:corrections3b} and \eqref{eq:corrections4}.

A time-dependent $\Delta\mathbf{g}$ can be due to small time-dependent contributions to the gravitational field, but also to fluctuations of the laser phase. Indeed, a time-dependent $\mathbf{g}'$, which involves a substitution analogous to Eq.~\eqref{eq:time_dependent} into Eq.~\eqref{eq:chirp4}, can entirely capture the effects of laser frequency noise and laser phase noise.
Both noise sources are closely related due to the differential relation between laser phase and frequency: $\omega_\text{chirp} = (d\varphi / dt)_\text{chirp}$.
They can be due to fluctuations of the laser source and to vibrations of the optical fibers or the retro-reflection mirror.
\comment{As discussed in Sec.~\ref{sec:atom_interferometer},} 
the phase-shift contributions due to time-dependent perturbations of the gravitational field can be cancelled out by simultaneously operating two reversed interferometers (with opposite sign for $\hat{\mathbf{n}}$).
On the other hand, the effects of laser phase noise can be suppressed by considering the differential phase shift for a gradiometer-like configuration involving two atom interferometers with different initial velocities, $\mathbf{v}_0^A$ and $\mathbf{v}_0^B$, interrogated by a common laser beam.
Due to the different initial velocities, two different laser frequencies are needed in order to address the two interferometers. These can be generated with an acousto-optical modulator (AOM) acting on a single-frequency carrier beam so that both frequency components undergo common propagation and any differential phase noise is minimized, as explained in Sec.~\ref{sec:implementation}.

It should also be noted that the finite duration of the laser pulses has not been explicitly considered in this Appendix. This point will be discussed in detail in a future publication, but the main conclusions are similar to those reached in Ref.~\cite{antoine06b} for two-photon transitions.
In particular, one can naturally regard the pulse times considered above as the central times for each pulse. Additional phase-shift corrections arise then due to the finite pulse duration $\tau$. The main ones have the same form as the first term on the right-hand side of Eq.~\eqref{eq:corrections4} but suppressed by 
a factor $(\tau / T)$ or $(\tau / T)^2$.
\comment{Furthermore, these contributions will cancel out when adding the phase shifts for reversed interferometers.}

\appsection{Arm trajectories}
\label{app:arm_trajectories}

In Appendix~\ref{app:phase-shift} the interferometric phase shift was obtained by making use of the proper times calculated along the mid-point trajectory.
Here we will analyze the differences that arise when considering the actual proper times along the central trajectories of the two interferometer arms and show that they imply no significant modifications of the phase-shift result at the order at which we are working.

\subsection{Recoil velocity and mid-point trajectory}
\label{sec:recoil_vel}

We work in the freely falling frame comoving with the mid-point trajectory, where the velocities of the two arm trajectories approximately amount to half the recoil velocity, $\pm \mathbf{v}_\text{rec} / 2$, which is more precisely defined as follows. An atom initially at rest and undergoing the clock transition from the excited to the ground state, emits a photon with momentum
\begin{equation}
\hbar\, \mathbf{k} = \frac{\Delta E}{c} \, \hat{\mathbf{n}} \,\,
\bigg[ \, 1 - \frac{\Delta E}{2\, m\,c^2} + O \Big( \big( \Delta E / m c^2 \big)^2 \Big) \, \bigg]
\label{eq:recoil_momentum} ,
\end{equation}
which takes into account the recoil energy of the atom after emitting the photon. Exactly the same momentum, but with opposite sign, is acquired by the emitting atom and corresponds to a recoil velocity $- \mathbf{v}_\text{rec}$ with
\begin{equation}
\mathbf{v}_\text{rec} = \frac{\hbar\, \mathbf{k}}{m} \,
\bigg[ \, 1 + O \Big( \big( \Delta E / m c^2 \big)^2 \Big) \, \bigg]
\label{eq:recoil_velocity} ,
\end{equation}
\comment{which is typically defined for the case of photon absorption rather than emission.}
Moreover, in the reference frame where the atom has the same velocity before and after the photon emission, but with opposite directions, these velocities are given by
\begin{equation}
\mathbf{v}_\pm = \,\pm\, \frac{\mathbf{v}_\text{rec}}{2}
\, \bigg[ \, 1 + O \Big( v_\text{rec}^2 / c^2 \Big) \, \bigg]
\label{eq:half_recoil_velocity} .
\end{equation}
because the relativistic corrections associated with the composition of velocities contribute at higher order.

On the other hand, as done in Fig.~\ref{fig:ff_frame}, it is actually convenient to consider the mid-point trajectory defined by the intersections of the two arm trajectories with light rays, i.e.\ such that for any light ray intersecting the two arm world lines the distance of the intersection points to the mid-point trajectory is the same. In the reference frame where the mid-point world line defined in this way is at rest the velocities of the two arms are slightly different and given by
\begin{equation}
\mathbf{v}'_\pm =  \,\pm\, \frac{\mathbf{v}_\text{rec}}{2}
\, \bigg[ \, 1 \mp \frac{v_\text{rec}}{2 c}
+ O \Big( v_\text{rec}^2 / c^2 \Big) \, \bigg]
\label{eq:null_half_recoil} .
\end{equation}
\comment{This definition of the mid-point world line avoids small sudden accelerations that would otherwise arise when the second laser pulse intersects the arm trajectories, so that it remains a spacetime geodesic (i.e.\ a freely falling trajectory) during the whole interferometer sequence.}

\subsection{Proper time along the arm trajectories}
\label{sec:arm_traj}

The propagation phase for an arm segment between the $j$th and $(j+1)$th laser pulses, which is proportional to the proper time along the corresponding central trajectory of the atomic wave packet, $\mathbf{X}_\text{c} (t'_\text{FW})$, can be written as follows in the Fermi-Walker frame associated with the mid-point trajectory:
\begin{align}
\mathcal{S}_n &= - m_n\, c^2  \int^{\tau^{(j+1)}}_{\tau^{(j)}} \!\!\! d\tau_\mathrm{c}
\nonumber \\
&= \int^{\tilde{t}_\text{FW}^{(j+1)}}_{\tilde{t}_\text{FW}^{(j)}} \!\!  dt'_\text{FW} \,
\bigg[ - m_n\, c^2 + \frac{1}{2}\, m_n \left( \frac{d \mathbf{X}_\text{c}}{d t'_\text{FW}} \right)^2
\nonumber \\
& \qquad\qquad\qquad\quad \,
- m_n\,  U_\text{FW} \big( t'_\text{FW},\mathbf{X}_\text{c} \big)  + O \Big( 1/c^2\Big) \, \bigg]
\label{eq:prop_phase_FW} ,
\end{align}
where $\tilde{t}_\text{FW}^{(j)}$ and $\tilde{t}_\text{FW}^{(j+1)}$ denote the Fermi-Walker times at which the central wave fronts of each pulse
intersect the arm trajectory.
The gravitational potential $U_\text{FW}$ vanishes for the case of a uniform 
field in the laboratory frame, whereas  in the presence of a gravity gradient it is given by $U_\text{FW} (\mathbf{X}_\text{c}) = - (1/2)\, \mathbf{X}_\text{c}^\mathrm{T} \, \Gamma \, \mathbf{X}_\text{c}$.

To calculate the proper time along the arm trajectories, 
one can then proceed analogously to the derivation in Appendix~\ref{app:ff_frame} of the relation between the proper time along the mid-point trajectory and the time in the laboratory frame.
Indeed, the time $t_\text{FW}$ for a light ray intersecting the mid-point world line is related to the time $\tilde{t}_\text{FW}$ when it intersects the central trajectory $\mathbf{X}_\text{c} (t'_\text{FW})$
through a relation analogous to Eq.~\eqref{eq:Doppler2}:
\begin{equation}
\quad \ 
\frac{d \tilde{t}_\text{FW}}{d t_\text{FW}} = \frac{1}{1 - \hat{\mathbf{n}} \cdot \mathbf{v}_\text{c} / c}
\label{eq:Doppler_FW} ,
\end{equation}
where $\mathbf{v}_\text{c} = d \mathbf{X}_\text{c} / d \tilde{t}_\text{FW}$ is the velocity of the central trajectory in the Fermi-Walker frame.
Similarly, the proper time $\tau_\text{c}$ along the central trajectory is connected to this intersection time $\tilde{t}_\text{FW}$ through a relation analogous to Eq.~\eqref{eq:dilation}:
\begin{equation}
\frac{d \tau_\text{c}}{d \tilde{t}_\text{FW}} \,=\, 1 - \frac{1}{2\, c^2}\! \left( \frac{d \mathbf{X}_\text{c}}{d \tilde{t}_\text{FW}} \right)^2
+\, \frac{1}{c^2}\, U_\text{FW} \big( \tilde{t}_\text{FW}, \mathbf{X}_\text{c} \big)
\label{eq:dilation_FW} .
\end{equation}

\subsubsection{Contributions proportional to $\Delta m$}

The contributions of order $1/c$ from Eq.~\eqref{eq:Doppler_FW} cancel out for an exactly symmetric Mach-Zehnder interferometer. On the other hand, the time asymmetries that arise for a non-vanishing $\Delta \mathbf{g}$ lead to terms proportional to $T \, ( \mathbf{v}_\text{rec} \cdot \Delta \mathbf{g} \, T) / c^2$, which can be neglected compared to the corrections in Eq.~\eqref{eq:corrections4} since one typically has $v_\text{rec} \lesssim 10^{-3}\, v_0$.
(Similar contributions result from the slightly modified momentum kick of the second pulse due to the term proportional to $\hat{\mathbf{n}} \cdot \Delta \mathbf{g} \, T / c$ in the Doppler shift of the laser wavelength.)
In contrast, the contributions of gravity gradients to $\mathbf{v}_\text{c}$ in Eq.~\eqref{eq:Doppler_FW} lead to terms of order $T \, \big( \mathbf{v}_\text{rec}\, ( \Gamma \, T^2 ) \, \hat{\mathbf{n}} \big) / c$, which would not be entirely negligible.
Nevertheless, one can show that, in fact, in all these cases there is
a vanishing net contribution of such terms to the phase shift. This can be proven as follows.

Let us focus on the first term in the integrand of Eq.~\eqref{eq:prop_phase_FW}, make use of Eq.~\eqref{eq:Doppler_FW} and expand the denominator in powers of $1/c$. The zeroth-order term simply corresponds to the proper time along the mid-point trajectory. On the other hand, the first-order term gives rise to the following contribution proportional to $\Delta m$ for an arm segment where the atoms are in the excited state:
\begin{equation}
- \frac{\Delta E}{\hbar}
\int^{\tilde{t}_\text{FW}^{(j+1)}}_{\tilde{t}_\text{FW}^{(j)}} \!\!\! d\tilde{t}_\text{FW} \,
\frac{\hat{\mathbf{n}} \cdot \mathbf{v}_\text{c}}{c}
\,=\, -\mathbf{k}' \cdot \mathbf{X}_\text{c} ^{(j+1)} + \mathbf{k}' \cdot \mathbf{X}_\text{c} ^{(j)}
\label{eq:eqivalence_FW} ,
\end{equation}
where the $j$th and $(j+1)$th pulses drive the transition from the ground to the excited state and vice versa.
Here $\mathbf{k}' = (\Delta E / \hbar c)\, \hat{\mathbf{n}}$ and have introduced the notation $\mathbf{X}_\text{c} ^{(l)} \equiv \mathbf{X}_\text{c} \big( \tilde{t}_\text{FW}^{(l)} \big)$ for each pulse.
In particular, $\mathbf{X}_\text{c} ^{(j)}$ vanishes for the first laser pulse in the interferometer sequence, whereas for the second one it has the same value but opposite sign for the two arms%
\footnote{\comment{For gravity gradients there is a slight difference connected with light's extra time of flight for the small displacement caused by tidal effects, but it is suppressed by an additional factor $v_\text{rec} / c$.}}.
Hence, their contributions to the phase shift cancel out. Finally, as discussed below in Appendix~\ref{app:open_interferometers}, either $\Delta \mathbf{g} \neq \mathbf{0}$ or gravity gradients 
lead to an open interferometer and a non-vanishing contribution of the term proportional to $\mathbf{X}_\text{c} ^{(j)}$ for the last pulse. However, as will be shown there, this contribution and the so-called separation phase that needs to be taken into account for open interferometers do cancel out.
\comment{Similar conclusions hold for small shifts in the pulse timing, considered in Appendix~\ref{sec:pulse_timings}.}


In the previous paragraph we have focused on the term of order $1/c$ that arises when expanding the right-hand side of Eq.~\eqref{eq:Doppler_FW}. Terms of order $(\hat{\mathbf{n}} \cdot \mathbf{v}_\text{c} / c)^2$ and higher are negligible. Indeed, for $\mathbf{v}_\text{c} = \pm \mathbf{v}_\text{rec} / 2$ the contribution from one arm before the central pulse coincides with that from the other arm after the pulse. This exact coincidence does not necessarily hold for perturbed trajectories due to $\Delta \mathbf{g}$ or gravity gradients and leading to $\mathbf{v}_\text{c} = ( \mathbf{v}_\text{rec} + \Delta\mathbf{v}_\text{rec} ) / 2 $, but the differences are of order $\big( v_\text{rec}^2 / c^2 \big) \big( \Delta v_\text{rec} / v_\text{rec} \big)$.
In the first case one has $\big( \Delta v_\text{rec} / v_\text{rec} \big) \sim \Delta g\, T / c$, which leads to a contribution of order $1/c^3$. For gravity gradients $\big( \Delta v_\text{rec} / v_\text{rec} \big) \sim \big( \Gamma\, T^2 \big)$ and their contribution can also be neglected because $v_\text{rec}^2 / c^2 \lesssim 10^{-6} \times v_0 (g\, T) / c^2$ and $\big( \Gamma\, T^2 \big) \sim 3 \times 10^{-6}$.
Note also that similar considerations to those made in this paragraph apply to the terms of order $1/c^2$ in Eq.~\eqref{eq:dilation_FW}.

\subsubsection{Contributions proportional to $m$}

So far we have discussed contributions proportional to $\Delta m$ (or, equivalently, $\Delta E$) and corresponding to arm segments with the atoms in the excited state. Let us now consider the contributions proportional to $m$, which arise for all arm segments. The sum of the contributions from the first term in the integrand of Eq.~\eqref{eq:prop_phase_FW} is the same along the two arms, so that there is not net contribution to the phase shift. On the other hand, contributions proportional to $m v_\text{rec}^2$ are not necessarily negligible because $m\, v_\text{rec} / c \sim \Delta m$.
Nevertheless, since the central trajectories for the two arms are typically symmetric with respect to the mid-point trajectory, a non-vanishing contribution to the phase shift can only arise during light's short time of flight between the two arms, of order $( v_\text{rec} / c )\, T$, and provided that there is a different $\Delta v_\text{rec}$ before and after the second pulse. This amounts to phase-shift contributions of order $m\, v_\text{rec}\, \Delta v_\text{rec} \, (v_\text{rec} / c)\, T \sim \Delta m \, v_\text{rec}\, \Delta v_\text{rec} \, T$, which can be neglected for both $\Delta v_\text{rec} \sim v_\text{rec} \, \big( \Gamma\, T^2 \big)$ and  $\Delta v_\text{rec} \sim \Delta g \, T$, as explained above.

Finally, note that gravity gradients lead to a slightly different Doppler shift of the laser wavelength (and the corresponding momentum kick) for each arm. Because of this asymmetry between the two arms, the contribution to the phase shift is not restricted to light's short time of flight 
and is instead of order $m\, v_\text{rec}\, \Delta v_\text{rec} \, T$. However, since $\Delta v_\text{rec} \sim v_\text{rec} \, (v_\text{rec} / c)\, \big( \Gamma\, T^2 \big)$  in this case, the contribution can be neglected as well.

\subsubsection{Total contribution}

From the above analysis for the terms proportional to $m$ and $\Delta m$ we can conclude that
the net phase-shift contributions that result from calculating the proper times along the arm trajectories rather than the mid-point world line are of higher order and can indeed be safely neglected.

\subsection{Local anharmonicities and equivalence to the two-photon diffraction case}
\label{sec:anharm}

The arguments in the previous subsection assumed uniform gravity gradients, which correspond to harmonic gravitational potentials, and need to be generalized for non-uniform ones.
Indeed, if local anharmonicities of the gravitational potential over length scales comparable to the arm separation are non-negligible, they can lead to asymmetries in the dynamics of the two interferometer arms and some of the arguments must be reconsidered.
In particular, if we choose the mid-point trajectory to be a freely falling one (i.e.\ a spacetime geodesic), these local anharmonicities will \comment{generally} lead to an asymmetric spatial separation with respect to the two arm trajectories.

Interestingly, in order to analyze the effects of such anharmonicities, one can take advantage of the close connection with the computation for conventional atom interferometers based on two-photon diffraction processes such as Raman or Bragg.
More specifically, the relevant contributions proportional to $\Delta m$ are captured by Eq.~\eqref{eq:eqivalence_FW} for each arm segment, which coincides with the phases acquired by an atomic wave packet when diffracted by Raman or Bragg pulses \comment{if one takes $\mathbf{k}'$ to be the wave vector associated with the two-photon momentum transfer.} Moreover, the terms proportional to $m$ in the propagation phase are the same for interferometers based on single- and two-photon transitions,
and similarly for the separation phase in case of an open interferometer.
Therefore, the effects of anharmonicities in the freely falling frame will be equivalent to those found in studies for atom interferometers based on two-photon transitions such as Ref.~\cite{overstreet21}.
\comment{[The effects of light's time of flight between the two arms for the terms proportional to $m$ is suppressed by an additional power of $v_\text{rec} / c$ and can be neglected. The same applies to the small difference (involving terms of higher-order in $v_\text{rec} / c$) between the wave vector $\mathbf{k}'$ appearing on the right-hand side of Eq.~\eqref{eq:eqivalence_FW} and the recoil momentum $\hbar\, \mathbf{k}$ in Eq.~\eqref{eq:recoil_momentum}.]}

The anharmonicities of the gravitational potential generated by a dense source mass placed close to the apex of an atom interferometer with a large arm separation played a key role in the atomic fountain experiments reported in Ref.~\cite{overstreet22}. In contrast, for the experimental implementation considered in Sec.~\ref{sec:implementation} the effect of local anharmonicities will be much smaller and will be suppressed when adding the phase shifts for reversed interferometers with opposite $\hat{\mathbf{n}}$.

Finally, it is worth mentioning that the considerations in this subsection also apply to state-dependent external potentials. The analysis in Appendix~\ref{app:external_forces} below explicitly considers external forces that are approximately uniform (i.e.\ approximately linear potentials) over length scales comparable to the arm separation. For state-dependent quadratic potentials, on the other hand, analogous effects to those discussed above for local anharmonicities will arise and a similar approach taking advantage of the equivalence with the two-photon case can be employed.

\appsection{Open interferometers}
\label{app:open_interferometers}

\highlight{
Various causes can lead to open atom interferometers with a relative displacement between the central trajectories of the atomic wave packets interfering at the exit ports
(either a relaitve displacement $\delta \mathbf{X}$ in position, a relative displacement $\delta \mathbf{P}$  in momentum or both).
For example, the tidal forces associated with gravity gradients give rise to relative displacements given by Eq.~\eqref{eq:displacements_gg} and, similarly, rotations originate displacements given by Eqs.~\eqref{eq:displacements_rot1}--\eqref{eq:displacements_rot2} below.
Importantly, open interferometers result in sensitivity of the interferometric signal to initial conditions
\cite{roura14,roura17a},
which can also imply a loss of contrast when inferring the phase shift from the total number of atoms detected at each exit port \cite{roura14}.

Furthermore, as explained for instance in Refs.~\cite{dimopoulos08a,roura20a},
besides the propagation phases along the two arms, for open interferometers there is an additional phase-shift contribution associated with the relative displacement between the interfering  wave packets, which is commonly known as \emph{separation phase} and is given by
\begin{equation}
\delta\phi_\text{sep} = - \bar{\mathbf{P}} \cdot \delta \mathbf{X} / \hbar
\label{eq:separation_phase} ,
\end{equation}
where $\bar{\mathbf{P}}$ is the average central momentum for the two interfering wave packets at the exit port.
Interestingly, in the freely falling frame where the mid-point trajectory is at rest this separation phase and the phase-shift contribution connected with the last beam-splitter cancel out at leading order in $1/c$.
\comment{Indeed, the central trajectories for the two arms when the last beam-plitter pulse is applied will be respectively displaced by $\pm\, \delta \mathbf{X} / 2$ with respect to the mid-point trajectory. Hence, for exit port I the time spent in the excited state by the atoms following the lower arm will be shifted by $- (\delta \mathbf{X} \cdot \hat{\mathbf{n}}) / 2 c$ to leading order in $1/c$, which results in a phase-shift contribution of $-(\Delta E / \hbar)\, (\delta \mathbf{X} \cdot \hat{\mathbf{n}}) / 2 c$.
This contribution cancels out the separation phase for that port, given by
$(m \mathbf{v}_\text{rec} / 2 \hbar) \cdot \delta \mathbf{X}$,
since $m \mathbf{v}_\text{rec} = (\Delta E / c)\, \hat{\mathbf{n}} \left( 1 + O \big( \Delta E / m c^2 \big) \right)$.
Analogous conclusions hold for exit port II.}

The effects of gravity gradients and rotations will be specifically considered in Appendix~\ref{app:gg_rotations}.
Here we will focus instead on the role of the frequency chirp and the pulse timings as possible causes of open interferometers, and on the associated phase-shift corrections.
}

\subsection{Frequency chirp}
\label{sec:frequency_chirp}

A non-vanishing acceleration in the Doppler factor, given by Eq.~\eqref{eq:Doppler2}, leads to an open interferometer.
This point can be straightforwardly seen in the freely falling frame, where the uncompensated time-dependent Doppler factor gives rise to different wave vectors (and associated momentum transfers) for the various laser pulses. It also implies a small time asymmetry between the first two pulses and the last two.
These two effects combined result in a relative displacement
\begin{equation}
\comment{\delta \mathbf{X} =  (\mathbf{v}_\text{rec} / c)\, ( \hat{\mathbf{n}} \cdot \Delta\mathbf{g} )\, T^2}
\end{equation}
between the interfering wave packets.
Fortunately, having such an open interferometer with undesirable consequences is avoided when compensating the Doppler factor through a suitable frequency chirp as discussed in 
Appendix~\ref{app:phase-shift}. Indeed, the exact compensation of the Doppler effect guarantees that the wave vectors for the various laser pulses remain the same in the freely falling frame and that no timing asymmetry is present.

These conclusions differ from those of Refs.~\cite{bott22a,bott22b}, where it was claimed that the frequency chirp needed for keeping single-photon transitions on resonance%
\footnote{The frequency detuning for a single-photon transition in a uniform gravitational field including the effects of a linear frequency chirp and $\Delta m$ has recently been calculated in Ref.~\cite{bott23}.}
for atoms falling in a gravitational field leads to an open interferometer. The analysis there was based on the resulting change of momentum transfer in the laboratory frame. However,  it overlooked the fact that the change of internal state involves a momentum change even for vanishing recoil velocity. Indeed, the total momentum change for an atom with velocity $\mathbf{v}$ is given by
\begin{equation}
\Delta \mathbf{p} = \Delta m\, \mathbf{v} + m\, \Delta \mathbf{v}
+ O \big( 1/c^2 \big)
\end{equation}
Therefore, for a fixed recoil velocity $\Delta \mathbf{v}$, the first term implies an additional momentum that depends on the velocity and will change for every pulse as the atoms fall in the gravitational field%
\footnote{In fact, this additional momentum is closely related to the residual recoil pointed out in Ref.~\cite{roura20a}.}.
The change of momentum transfer associated with the frequency chirp actually coincides with this additional momentum and guarantees that the recoil velocity is the same for all laser pulses, so that the interferometer remains closed.

In particular, this means that for perfect compensation (corresponding to $\Delta  \mathbf{g} = \mathbf{0}$) and fixed $\mathbf{v}'_0$, there is no sensitivity of the phase shift to small changes of the atomic wave packet's initial velocity. Indeed, one can immediately see that the dependence on $\mathbf{v}_0$ of the right-hand side of Eq.~\eqref{eq:phase_shift2} and of the term proportional to $(\hat{\mathbf{n}} \cdot \Delta \bar{\mathbf{v}}_0)$
in Eq.~\eqref{eq:corrections4} cancel out exactly.
On the other hand, for an imperfect cancelation of the Doppler factor there will be a small relative displacement proportional to $\Delta \mathbf{g}$ between the interfering wave packets, but the resulting sensitivity to the initial velocity will be suppressed by $\Delta g / g \lesssim 10^{-5}$.

In addition, there is an even smaller effect connected with the impact of the time dilation factor in Eq.~\eqref{eq:dilation} on the wave vectors of the laser pulses and the timings between them.
However, they give rise to much smaller relative displacements between the interfering wave packets, and to associated phase-shift contributions of order $1/c^3$, which are 
negligible.

\subsection{Pulse timings}
\label{sec:pulse_timings}

So far we have assumed that laser phase and pulse timing are linked and equally affected by any perturbations. This should indeed be the case for the Doppler shift and for vibrations of the retro-reflection mirror or the optical fibers injecting the laser beams. However, imperfections in the pulse generation as far as their envelope is concerned, \highlight{incomplete cancelation of the Doppler effect on the pulse timing \comment{or any source of laser phase noise before the pulse generation can} all} lead to shifts of the pulse central time with respect to the laser wave fronts, whose implications will be analyzed here.

Let us consider first the case of perfect cancellation of the Doppler factor through the frequency chirp and no phase noise. A small shift by $\Delta T$ of the second laser pulse while keeping the timings%
\footnote{\comment{The times $T$ and $T + \Delta T$ correspond to the emission times if no chirping were applied and coincide with the time coordinate $\bar{t}$ in case of perfect cancellation of the Doppler factor.}}
of the first and third pulses at $0$ and $2 T$, implies the replacement $T \to T + \Delta T$ for the intermediate time appearing in the integration limits on the right-hand side of Eq.~\eqref{eq:phase_shift1} and leads to a phase-shift correction approximately given by $- 2\, (\Delta E / \hbar)\, \Delta T$. Moreover, in this case there is also a non-vanishing net contribution from the laser phases. Indeed, when calculating the total propagation phase along each interferometer arm, each laser pulse contributes with a \comment{phase factor $\exp (i\, \varepsilon_j\, \varphi_j)$, where $\varphi_j = \int^{t_j} (d\varphi / dt)_\text{chirp}\, dt$ and $t_j$ is the central time for that pulse;}
$\varepsilon_j = -1, 0, 1$ depending on whether a photon is absorbed, there is no transition or a photon is emitted.
\highlight{In a Mach-Zehnder interferometer the contribution of these phases to the interferometer phase shift is given by $\delta\varphi = - \varphi_3 + 2\, \varphi_2 - \varphi_1$. For perfect pulse timings
one has $\varphi_2 - \varphi_1 = \varphi_3 - \varphi_2 = \omega_0\, T$,
which leads to a vanishing net contribution%
\footnote{\highlight{This might not be the case if the generation of each pulse envelope gave rise to an additional pulse-dependent contribution to $\varphi_j$, but such a non-vanishing contribution would still cancel out in the differential phase shift for the gradiometric configuration.}}
$\delta\varphi = 0$.
However, this is no longer the case in the presence of a time asymmetry $\Delta T$ leading to $\varphi_2 - \varphi_1 = \omega_0 \, (T + \Delta T)$, $\varphi_3 - \varphi_2 = \omega_0 \, (T - \Delta T)$ and resulting in $\delta\varphi = 2\, \omega_0 \, \Delta T$.}
When combined with the above result for the modified Eq.~\eqref{eq:phase_shift1}, one finally gets the following phase-shift correction due to the imperfect pulse timing:
\begin{equation}
\delta\phi_\text{timing} = - 2 \left( \frac{\Delta E}{\hbar} - \omega_0 \right) \Delta T = - 2\, \delta \, \Delta T
\label{eq:pulse_timing1} ,
\end{equation}
which vanishes for a vanishing detuning $\delta = \Delta E / \hbar - \omega_0$.

In general, there will only be an incomplete cancellation of the Doppler factor and one will also need to make the replacement $T \to T + \Delta T$ for the intermediate time in the integration limits of Eq.~\eqref{eq:corrections3a}. The resulting extra terms can be taken into account with the following redefinition of the detuning $\delta$ in Eq.~\eqref{eq:pulse_timing1}:
\comment{
\begin{equation}
\delta = \frac{\Delta E}{\hbar}
\left[ 1 + \frac{ ( \hat{\mathbf{n}} \cdot \Delta \bar{\mathbf{v}}_0 ) }{c}
+\frac{ ( \hat{\mathbf{n}} \cdot \Delta \mathbf{g} ) }{c}\, T \right]
 - \omega_0
\label{eq:pulse_timing2} .
\end{equation}
}%
\comment{where terms of order $1/c^2$ connected with the time-dilation factor $(d \bar{\tau} / d \bar{t})$ have been omitted.}
As explained in Appendix~\ref{sec:phase_noise}, the result 
can be straightforwardly generalized to the case of a time-dependent $\Delta\mathbf{g}$, which can also account for laser phase noise.

Shifting the second pulse by $\Delta T$ implies an asymmetry in the time separation
with the two beam-splitter pulses and results in an open interferometer with relative displacement $\delta \mathbf{X} = 2\, \mathbf{v}_\text{rec} \, \Delta T$. The expected phase-shift dependence on the initial velocity associated with such a relative displacement \cite{roura14,roura17a} agrees with the dependence of $\delta\phi_\text{timing}$ on $\Delta \bar{\mathbf{v}}_0$ when substituting Eq.~\eqref{eq:pulse_timing2} into the right-hand side of Eq.~\eqref{eq:pulse_timing1}.
 \comment{Moreover, as explained 
above in the second paragraph of this appendix, for open interferometers there is an additional phase-shift contribution known as separation phase, but in the freely falling frame comoving with the mid-point trajectory it is canceled out by the phase contribution connected with the last beam-splitter pulse, as explained above.}

It should be noted that changes of the first and third laser pulses that keep an equal time separation $T + \Delta T$ with the second pulse can be easily taken into account by simply replacing $T$ with $T + \Delta T$ in any of the results for the regular Mach-Zehnder interferometer such as
\pagebreak[4]
Eqs.~\eqref{eq:phase_shift2} and \eqref{eq:corrections4}. Furthermore, an arbitrary change of the timings for the three laser pulses can always be reduced to a combination of this case and the timing asymmetry for the second pulse discussed above.

We close this appendix with a discussion of the requirements on pulse timing errors and their impact on the proposed measurements. A detuning $\delta = 2\pi \times 300\, \text{Hz}$ and a timing asymmetry $\Delta T = 1\, \mu\text{s}$ lead to a phase-shift correction $\delta\phi_\text{timing} \sim 4\, \text{mrad}$, which allows a time-dilation measurement at the $10^{-4}$ level with the experimental implementation considered in Sec.~\ref{sec:implementation}. Moreover, reducing $\delta$ and $\Delta T$ so that their product decreases by one or two orders of magnitude and brings the associated uncertainty down to the $10^{-5}$ or even  $10^{-6}$ level seems quite feasible.
\highlight{(By comparison, a completely uncompensated Doppler effect on the pulse timing would correspond to $\Delta T \sim 0.1\, \mu\text{s}$ for $T = 1\, \text{s}$.)
On the other hand,} a symmetric change by $\Delta T$ of the pulse separations
also leads to a relative change of $2\, \Delta T / T$ to the phase-shift signal in Eq.~\eqref{eq:phase_shift2} and amounts to $\Delta T / T \sim 10^{-6}$ for $\Delta T = 1\, \mu\text{s}$ and $T = 1\, \text{s}$.
More importantly, if there is a slightly different $\Delta T$ for the two reversed interferometers, there will not be a complete cancellation of the correction terms linear in $\hat{\mathbf{n}}$.
For example, for a difference between the pair of reversed interferometers corresponding to $\Delta T = 0.1\, \mu\text{s}$, the remaining contribution from the first term on the right-hand side of Eq.~\eqref{eq:corrections1} after partial cancellation will be comparable to 
the term proportional to $( \bar{\mathbf{v}}_0 \cdot \Delta\mathbf{g} / c^2 )\, T^2$.

In fact, for the gradiometric configuration displayed in Fig.~\ref{fig:gradiometric_conf} having a slightly different pulse separation in the comoving frame of each interferometer is unavoidable if one uses a common pulse envelope for the frequency components addressing each one of the two interferometers. Indeed, due to their different initial velocities the Doppler-shift contribution $\Delta T = ( \hat{\mathbf{n}} \cdot \Delta\mathbf{v}_0 / c )\, T$ leads to different pulse separations for interferometers $A$ and $B$. Moreover, this also implies that it is not possible to have the same pulse separation for each pair of reversed interferometers: if one adjusts the pulse emission times so that they are the same for interferometer $A$ and the reversed one, this will not be the case for interferometer $B$ and its reversed counterpart. The resulting incomplete cancellation of the contributions from the first term on the right-hand side of Eq.~\eqref{eq:corrections1} for the two 
interferometers with opposite $\hat{\mathbf{n}}$ gives rise to a phase-shift correction that coincides, up to a coefficient of order one, with the term proportional to $( \bar{\mathbf{v}}_0 \cdot \Delta\mathbf{g} / c^2 )\, T^2$.
Similarly, for the gravity gradient corrections in Eq.~\eqref{eq:phase_shift_gg} below and $v_0 = 20\, \text{m/s}$ such a change of the pulse separation gives rise to a contribution comparable to that for $\Delta \mathbf{g}$ \comment{with $\Delta g \sim 10^{-5} g$.}
In addition, in the configuration with upward-pointing $\hat{\mathbf{n}}$, displayed in Fig.~\ref{fig:gradiometric_conf}b), light's time of flight to the retro-reflecting mirror and back shifts the time of the first pulse for interferometer $A$  by $\Delta T \sim 2 L / c$. This shift changes the initial position $\bar{\mathbf{X}}_0$ in Eq.~\eqref{eq:phase_shift_gg} by $\bar{\mathbf{v}}_0\, \Delta T$ compared to the reversed interferometer in Fig~\ref{fig:gradiometric_conf}a and since we have roughly $L \sim 2\, v_0\, T$, it leads to a phase-shift contribution comparable to that due to the change of pulse separation.

\appsection{Gravity gradients and rotations}
\label{app:gg_rotations}

\subsection{Gravity gradients}
\label{sec:gravity_gradients}

Gravity gradients lead to open interferometers and phase-shift sensitivity to initial conditions.
Their leading contribution to the phase shift $\delta\phi$ can be obtained by considering the mid-point trajectory $\bar{\mathbf{X}} (\bar{t})$ given by Eq.~\eqref{eq:trajectory_gg} and substituting its time derivative $\bar{\mathbf{v}} = d \bar{\mathbf{X}} / d \bar{t}$ into Eq.~\eqref{eq:Doppler2}. This gives rise to additional terms on the right-hand side of Eq.~\eqref{eq:corrections2} and we will focus on the leading contributions, which are of order $1/c$ and linear in the \comment{gravity-gradient tensor $\Gamma$.} After substituting those terms into Eq.~\eqref{eq:corrections3a}, we obtain the following result for the leading correction due to gravity gradients:
\begin{equation}
\delta\phi_\text{gg} = \left( \frac{\Delta E}{\hbar\, c} \right) \,
\hat{\mathbf{n}}^\mathrm{T} \left( \Gamma \, T^2 \right)
\left[ \bar{\mathbf{X}}_0 + \bar{\mathbf{v}}_0 T
+ \frac{7}{12}\, \mathbf{g}\, T^2 \right]
\label{eq:phase_shift_gg} .
\end{equation}
The more common expression in terms of the initial position and velocity of the atomic wave packet,
$\mathbf{X}_0$ and $\mathbf{v}_0$,
is recovered by taking into account that $\bar{\mathbf{X}}_0 =  \mathbf{X}_0$ and $\bar{\mathbf{v}}_0 = \mathbf{v}_0 + \mathbf{v}_\mathrm{rec} / 2$. The latter relation clearly shows that when considering the reversed interferometer, 
one needs to change the initial velocity by $\mathbf{v}_\mathrm{rec}$, i.e.~$\mathbf{v}_0 \to \mathbf{v}_0 + \mathbf{v}_\mathrm{rec}$, so that $\bar{\mathbf{v}}_0$ and the mid-point trajectory remain unchanged.

\highlight{
In addition to their effect on the mid-point trajectory, gravity gradients also modify the central trajectories of the interferometer arms, which are no longer straight lines in the freely falling frame and do not overlap at the exit ports, resulting in an open interferometer as discussed in Appendix~\ref{sec:non-uniform_field}.
The differences that arise when calculating the propagation phase along these trajectories, compared to the evaluation along the mid-point world line, can be neglected once the phase-shift contribution from the separation phase is included, as explained in Appendix~\ref{sec:arm_traj}.
Therefore, the leading 
correction due to a uniform gravity gradient is entirely given by Eq.~\eqref{eq:phase_shift_gg}.

The generalization to non-uniform gravity gradients is relatively straightforward. Indeed, for a more general gravitational potential $U(t,\mathbf{X})$ one needs to find first the classical solution corresponding to the mid-point trajectory $\bar{\mathbf{X}} (\bar{t})$, which can be done perturbatively or by any other suitable method. The leading contribution, of order $1/c$, can then be directly obtained by making use of Eq.~\eqref{eq:eqivalence} derived below. As long as the gravitational potential around $\bar{\mathbf{X}} (\bar{t})$ can be locally approximated by a harmonic potential for length scales comparable to the arm separation, the gravitational field in the Fermi-Walker frame comoving with $\bar{\mathbf{X}} (\bar{t})$ can be characterized by the gravity gradient tensor $\Gamma_{ij} = -\partial^2 U / \partial x^i \partial x^j |_{\bar{\mathbf{X}} (\bar{t})}$. In this case the tensor $\Gamma ( t_\text{FW} )$ is 
time dependent, but the conclusions of Appendix~\ref{sec:arm_traj} still hold.
On the other hand, for locally anharmonic potentials, one needs to follow the approach of Appendix~\ref{sec:anharm}. Nevertheless, for the experimental implementation considered in Sec.~\ref{sec:implementation} the effect of such local anharmonicities will be quite small and will be suppressed when adding the phase shifts for the pair of interferometers with opposite $\hat{\mathbf{n}}$.
}

\subsection{Equivalence to two-photon transitions}
\label{sec:equivalence}

The phase-shift contribution in Eq.~\eqref{eq:phase_shift_gg} coincides with the result for light-pulse atom interferometers based on two-photon transitions such as Raman or Bragg diffraction. In fact, this equivalence holds more generally and can be understood as follows.
If one expands the Doppler factor in Eq.~\eqref{eq:Doppler2} in powers of $1/c$ and substitutes the leading contribution, given by $\hat{\mathbf{n}} \cdot \bar{\mathbf{v}} / c$, into Eq.~\eqref{eq:corrections3a}, one is left with terms of the following form:
\begin{equation}
- \frac{\Delta E}{\hbar}
\int^{\bar{t}_{j+1}}_{\bar{t}_j} \frac{\hat{\mathbf{n}} \cdot \bar{\mathbf{v}}}{c} \, d\bar{t}
\,=\, -\mathbf{k}'_{j+1} \cdot \bar{\mathbf{X}}_{j+1} + \mathbf{k}'_j \cdot \bar{\mathbf{X}}_j
\label{eq:eqivalence}\, ,
\end{equation}
where $\bar{t}_{j+1}$ and $\bar{t}_{j}$ are the times of the $j$th and $(j+1)$th \linebreak[4]
pulses, which drive the transitions to the excited state and back to the ground state respectively.
\highlight{In the equality we have introduced $\mathbf{k}'_j = k'\, \hat{\mathbf{n}}_j$ with $k' = \Delta E / (\hbar c)$, \comment{which coincide with the laser wave vectors to leading order in $1/c$.}}
For the second interferometer arm there is an analogous contribution but with opposite sign, which corresponds to the second term on the right-hand side of Eq.~\eqref{eq:corrections3a}.
After combining the contributions from both arms, the phase-shift result coincides with the so-called mid-point theorem \cite{borde02,antoine03b} for atom interferometers based on two-photon transitions, but in this case with single-photon rather than two-photon momentum transfers.

Furthermore, although Eq.~\eqref{eq:Doppler2} and the above derivation of Eq.~\eqref{eq:eqivalence} are specialized to the case where the direction $\hat{\mathbf{n}}_j$ is the same for all laser pulses, the result on the right-hand side of Eq.~\eqref{eq:Doppler2} is also valid in general. This point can be seen by considering a time-dependent rotation to a frame where the directions of all laser pulse are aligned, transforming back to the original frame at the end of the calculation and taking into account that the scalar product is invariant under rotations.

\subsection{Rotations}
\label{sec:rotations}

In particular, the considerations in the previous subsection
can be \comment{applied to determining} the effects of rotations on the interferometer phase shift. The leading correction due to Earth's rotation for the Mach-Zehnder interferometer considered \comment{in the present paper} is then given by
\begin{equation}
\delta\phi_\text{rot} = \comment{2} \left( \frac{\Delta E}{\hbar\, c} \right) 
\big[ \hat{\mathbf{n}} \cdot (\bar{\mathbf{v}}_0 \times \boldsymbol{\Omega})\, T^2
\comment{-\,} \hat{\mathbf{n}} \cdot (\mathbf{g} \times \boldsymbol{\Omega})\, T^3
\big]
\label{eq:phase_shift_rotation} ,
\end{equation}
where both the laser beam direction $\hat{\mathbf{n}}$ and the gravitational acceleration $\mathbf{g}$ are fixed in Earth's co-rotating frame, whose angular velocity is $\boldsymbol{\Omega}$.

\highlight{
Moreover, rotations give rise to open interferometers with the following relative displacement between the two interfering wave packets at each exit port:
\begin{subequations}
\begin{align}
\delta \boldsymbol{\mathbf{X}}
&= 2\, (\mathbf{v}_\text{rec}\, T) \times(\boldsymbol{\Omega}\, T)
+ O \big( (\Omega\, T)^2 \big) ,
\label{eq:displacements_rot1} \\
\delta \boldsymbol{\mathbf{P}} &
= O \big( (\Omega\, T)^2 \big) ,
\label{eq:displacements_rot2}
\end{align}
\end{subequations}
which can be understood as a consequence of the changing $\hat{\mathbf{n}}$ for the laser pulses in the inertial (non-rotating) frame or, alternatively, as a consequence of the Coriolis acceleration experienced by the atoms in the co-rotating frame \cite{antoine03b,kleinert15}.
The phase-shift dependence on initial conditions linked to these relative displacement \cite{roura14} agrees with the dependence on the initial velocity $\mathbf{v}_0$ in Eq.~\eqref{eq:phase_shift_rotation}.
}

\appsection{Violations of the equivalence principle}
\label{app:violations}

Following Ref.~\cite{roura20a}, where further details can be found, we will consider
dilaton models \cite{damour12,damour10a} as a consistent framework for investigating violations of Einstein's equivalence principle.
For non-relativistic velocities and weak gravitational fields, the effect of the dilaton field on the dynamics of test masses can be captured by considering the following replacement in Eq.~\eqref{eq:prop_phase2}:
\begin{equation}
m_n\,  U (t',\mathbf{X}) \, \rightarrow \, m_n\, (1 + \beta_n) \, U (t',\mathbf{X})
\label{eq:dilaton_replacement} ,
\end{equation}
where the parameter $\beta_n$ depends on the atomic species and also on the internal state.
In general, the parameter is different for different species and states, which implies a violation of the universality of free fall (UFF).
Furthermore, based on energy conservation arguments (and local Lorentz invariance), it has been shown  that violations of universality of gravitational redshift (UGR) follow from these violations of UFF \cite{nordtvedt75,wolf16,giulini11}.
For a two-level atom, such violations of UGR are characterized by the parameter
\begin{equation}
\alpha_\text{e-g} = (\beta_2 - \beta_1) \left( \frac{m}{\Delta m} \right)
\label{eq:ugr_violation} .
\end{equation}

Let us see what the implications 
are for our proposed interferometry scheme. First of all, it is convenient to define the semisum and difference parameters, $ \bar{\beta} = (\beta_1 + \beta_2) / 2$ and $\Delta\beta = \beta_2 - \beta_1$, for atoms in the two internal states.
In the laboratory frame the mid-point trajectory corresponds in this case to the trajectory of an object falling with the mean gravitational acceleration $\big(1 + \bar{\beta} \big)\, \mathbf{g}$.
On the other hand, in the Fermi-Walker 
frame comoving with this mid-point world line the two clock states experience small opposite accelerations: $- \Delta\beta\, \mathbf{g} / 2$ and $\Delta\beta\, \mathbf{g} / 2$ for the ground and excited state respectively.
The resulting effect on the central trajectories of the two arms resembles that of the tidal forces due to gravity gradients.

\highlight{
When calculating in this Fermi-Walker frame the propagation phases for each segment along the two interferometer arms,
the expression for the propagation phase can be written analogously to the right-hand side of Eq.~\eqref{eq:prop_phase_FW} but with the replacement
\begin{equation}
U_\text{FW} \big( t'_\text{FW},\mathbf{X}_\text{c} \big) \, \to \,
\varepsilon_n \frac{\Delta\beta}{2} \, \mathbf{g} \cdot \mathbf{X}_\text{c}
\label{eq:U_FW_replacement} ,
\end{equation}
\comment{as well as $m_n c^2 \to m_n c^2 \, \big(1 \, + \, \varepsilon_n\, (\Delta\beta / 2)\, U (t',\bar{\mathbf{X}}) / c^2 \big)$ for the first term in the integrand, and where $\varepsilon_n = \pm 1$ depending on the internal state.} Moreover, one can use the same kind of arguments as in Appendix~\ref{sec:arm_traj} to show that it is sufficient to calculate the propagation phases along the mid-point world line. In particular, the opposite accelerations experienced by the two clock states lead to an open interferometer and the separation phase that arises in that case cancels out the contribution of the Doppler factor to leading order in $1/c$,
as explained in Appendix~\ref{app:open_interferometers}.
In addition, contributions proportional to $\Delta\beta$ and suppressed by factors $\Delta m / m$ or $v_\text{rec} / c$ compared to the corrections appearing in Eq.~\eqref{eq:phase_shift_dilaton} below will be neglected. 
\comment{Likewise, subleading corrections proportional to $\bar{\beta}\, (g / c)\, (v_\text{rec} / c)^2 \, T^2$ that arise because in this frame light rays do not exactly correspond to straight world lines are also negligible.}
}

One is therefore left with the contribution that corresponds to evaluating the action along the mid-point trajectory with mean acceleration $\bar{\mathbf{g}}$ and with the replacement in Eq.~\eqref{eq:dilaton_replacement}, which gives the following result for the phase shift:
\begin{equation}
\delta\phi 
= - 2 \,(\Delta E / \hbar) \, \big( 1 + \alpha_\text{e-g} / 2 \big)
\left( \bar{\mathbf{v}}_0 \cdot \bar{\mathbf{g}}\, T^2 + \bar{g}^2 T^3  \right) / c^2
\label{eq:phase_shift_dilaton} ,
\end{equation}
where we have introduced the parameter \comment{$\alpha_\text{e-g}$}
defined above, which characterizes small violations of the UGR.
Note that the contribution proportional to \comment{$\alpha_\text{e-g}$} is multiplied by a factor $1/2$.
\comment{The reason} is that the terms proportional to $\Delta\beta$ come entirely from evaluating the gravitational-potential part of the action, which corresponds to the gravitational redshift and contributes exactly to half of the time dilation effect in Eqs.~\eqref{eq:ff-clock2_phase} and \eqref{eq:phase_shift2}. The other half, which corresponds to the special relativistic time dilation, comes from the kinetic term in the action and is independent of $\Delta\beta$.

The result in Eq.~\eqref{eq:phase_shift_dilaton} agrees with what one would find for a localized clock following a trajectory that coincides with the interferometer's mid-point trajectory
\comment{(typically thanks to a suitable guiding potential)}
and shows that the proposed atom interferometric measurements can also be employed to test the UGR.
\highlight{A similar result would be obtained for an atom interferometer based on Raman transitions between two hyperfine states, despite some differences connected with the fact that two-photon processes rather than single-photon transitions are involved  in that case. Since $\Delta E$ is five orders of magnitude smaller for hyperfine states, the measurement of relativistic time-dilation effects lies beyond the sensitivity of such an interferometer, but one can still place some bounds on the violation parameter $\alpha_\text{e-g}$.
In fact, in this context it becomes clear that the experimental results reported in Ref.~\cite{xu22} should actually be interpreted as a test of UGR rather than a test of UFF. Nevertheless, tests of UGR placing tighter bounds on the same parameters can be achieved by comparing atomic-fountain clocks employing the same atomic species and located at sufficiently different heights.}

\comment{It should be noted, on the other hand, that when $(m / \Delta m)\, \Delta\beta$ is comparable to or smaller than $\bar{\beta}$, the terms proportional to $\bar{\beta}\, \mathbf{g}$ in Eq.~\eqref{eq:phase_shift_dilaton} can be equally relevant and even become the dominant contribution associated with violations of the equivalence principle.
These contributions can be interpreted as a violation of UFF, but such violations are much more strongly constrained by conventional Mach-Zehnder interferometers based on Bragg (or Raman) diffraction, where they are not suppressed by the small factors $v_0 / c$ or $g\, T / c$ of order $10^{-8}$.}

\highlight{
Note also that if the dilaton field couples to the electromagnetic field, the propagation of electromagnetic waves is slightly modified and light rays are no longer null geodesics, \comment{but this effect is very small for wavelengths much shorter than Earth's radius}. 
Moreover, for a time-independent dilaton-field configuration, such as that sourced by Earth, the time-dilation result remains unchanged \comment{anyway} since it is still guaranteed by time-translation invariance for the electromagnetic wave fronts.
\comment{This point holds for both a localized clock and the atom interferometric scheme considered here.}
}

\appsection{External forces}
\label{app:external_forces}

A detailed analysis of relativistic wave-packet propagation in the presence of external forces and guiding potentials was provided in Ref.~\cite{roura20a}. Here we will focus on weak forces such as those due to residual magnetic fields and black-body radiation.

The interaction of a neutral atom with residual magnetic fields through its magnetic dipole moment or with far-detuned electromagnetic radiation can be described in terms of a state-dependent external potential $V_n (t, \mathbf{x})$.
Similarly to what we have done in the previous Appendix, the effect of the external potential can be taken into account by making the following replacement in Eq.~\eqref{eq:prop_phase2}:
\begin{equation}
U (t',\mathbf{X}) \, \rightarrow \, U (t',\mathbf{X}) + \frac{1}{m_n} V_n (t',\mathbf{X})
\label{eq:external_potential_replacement} .
\end{equation}
In addition, it is convenient to introduce the following linear combinations of the potentials for the two clock states:
\begin{equation}
\bar{V}_n \equiv \frac{m_n}{2} \left( \frac{V_1}{m_1} + \frac{V_2}{m_2} \right)
\label{eq:external_potential1} ,
\end{equation}
\begin{equation}
\delta V_n \equiv m_n \left( \frac{V_2}{m_2} - \frac{V_1}{m_1} \right)
\label{eq:external_potential2} ,
\end{equation}
where the ratios $\bar{V}_n / m_n$ and $\delta V_n / m_n$ are independent of the internal state, labeled by the subindex $n$.
Provided that the potentials are approximately linear over length scales comparable to the arm separation,
one can proceed analogously to 
Appendix~\ref{app:violations}. In particular, when determining the mid-point trajectory with respect to the laboratory frame, one needs to add to the gravitational acceleration $\mathbf{g}$
the mean acceleration associated with the external potential, which results in the total acceleration
$\bar{\mathbf{a}} = \mathbf{g} - \boldsymbol{\nabla} \, \bar{V}_n / m_n$.
Moreover, in the Fermi-Walker frame comoving with the mid-point world line obtained in this way, atoms in the two clock   states, and hence in the two arms of the Mach-Zehnder interferometer, experience opposite accelerations $\pm \delta \mathbf{a} / 2$ with $\delta \mathbf{a} = - \boldsymbol{\nabla} (\delta V_n) / m_n$, which leads in general to an open interferometer.

\highlight{
Making use of the results for the propagation of matter-wave packets in the presence of external forces derived in Appendix~B of Ref.~\cite{roura20a}, one can proceed analogously to what was done for dilaton models in Appendix~\ref{app:violations} and show that the main contribution to the phase shift corresponds to evaluating the action along the mid-point trajectory with mean acceleration $\bar{\mathbf{a}}$ and with the replacement in Eq.~\eqref{eq:external_potential_replacement}.
}%
Before doing so, it is convenient to write the contributions of the external potential in terms of $\bar{V}_n / m_n$ and $\delta V_n / m_n$. Furthermore, between any pair of consecutive laser pulses we have different internal states for the two arms of a Mach-Zehnder interferometer and, hence, the external potentials for the two states always contribute with opposite sign. It is therefore sufficient to consider the following linear combination:
\begin{equation}
V_2 (t',\bar{\mathbf{X}}) - V_1 (t',\bar{\mathbf{X}}) = \Delta m \, \frac{\bar{V}_n (t',\bar{\mathbf{X}})}{m_n}
+ \bar{m} \, \frac{\delta V_n (t',\bar{\mathbf{X}})}{m_n}
\label{eq:external_potential_diff} ,
\end{equation}
where $\bar{m} = (m_1 + m_2) / 2 \approx m$ and the ratios $\bar{V}_n / m_n$ and $\delta V_n / m_n$ are independent of the internal state, 
\comment{as pointed out above.}

The phase shift $\delta\phi$ can then be obtained by evaluating along the mid-point trajectory the difference of the actions $\mathcal{S}_n$ for the two arms with the replacement in Eq.~\eqref{eq:external_potential_replacement} and making use of Eq.~\eqref{eq:external_potential_diff}.
For simplicity of the final expressions we will consider here the example of time-independent potentials with linear spatial dependence, but one can easily extend the calculation to time-dependent potentials and general spatial dependence (but approximately linear for length scales comparable to the arm separation).
The result for the time-independent case is given by
\begin{align}
\delta\phi =& - \frac{\Delta E}{\hbar} \, \bigg[ \,
2 \left( \bar{\mathbf{v}}_0 \cdot \bar{\mathbf{a}}\, T^2 + \bar{\mathbf{a}}^2\, T^3  \right) / c^2
\nonumber \\
& \qquad\qquad + \left( \frac{m}{\Delta m} \right) \left( \delta \mathbf{a} \cdot \bar{\mathbf{v}}_0 \,T^2
+ \delta \mathbf{a} \cdot \bar{\mathbf{a}} \, T^3  \right) / c^2 \bigg]
\label{eq:phase_shift_external} ,
\end{align}
which has the same kind of structure as the result in Eq.~\eqref{eq:phase_shift_dilaton}.
In fact, one can recover the result for dilaton models obtained in Appendix~\ref{app:violations} as a particular case with $V_n (t',\mathbf{X}) = m_n \, \beta_n \, U (t',\mathbf{X})$. It is thus clear that external forces can mimic a violation of Einstein's equivalence principle.

The terms proportional to $\delta \mathbf{a}$, which are enhanced by a factor $(m / \Delta m) \sim 10^{11}$, will typically dominate over the other terms in Eq.~\eqref{eq:phase_shift_external} unless $\delta V_n$ is much smaller than $\bar{V}_n$ by a factor of order $\Delta m / m$ or smaller.
\comment{In this respect, it should also be noted that one can typically use the approximations $\bar{V}_n \approx (V_1 + V_2) / 2$ and $\delta V_n \approx (V_2 - V_1)$ when determining $\bar{V}_n$ and $\delta V_n$, but should instead employ $\delta V_n \approx (V_2 - V_1) - (\Delta m / m)\, \bar{V}_n$ if $| V_2 - V_1 | \lesssim (\Delta m / m)\, | \bar{V}_n |$.
Similar considerations would apply to $\bar{V}_n$ if $| V_1 + V_2 | \lesssim (\Delta m / m)\, | \delta V_n |$ due to (nearly) identical potentials $V_1$ and $V_2$ but with opposite sign.}

\highlight{Finally, note that the \comment{potential gradient} $\boldsymbol{\nabla} \, \bar{V}_n$ will also contribute to $\delta\phi_\text{corr}$ through the replacement $\mathbf{g} \to \bar{\mathbf{a}}$ in Eq.~\eqref{eq:corrections4}, but this does not imply any essential changes. In particular, the linear term proportional to $(\hat{\mathbf{n}} \cdot \Delta\bar{\mathbf{a}})$, which gives the dominant contribution, will still cancel out when adding up the phase shifts for the two reversed interferometers with opposite $\hat{\mathbf{n}}$.}

\subsection{Magnetic fields}
\label{sec:magnetic_fields}

Details about the Zeeman shifts for $^{87}\mathrm{Sr}$ can be found in Ref.~\cite{boyd07} and the supplemental material of Ref.~\cite{zheng22}. The \emph{linear} Zeeman shifts for the clock states are of the order of $1 \, \text{kHz/G}$. Such contributions to $\bar{V}_n$ give rise to a mean acceleration
\begin{equation}
\bar{a} \sim 4 \times 10^{-6}\, \text{m/s}^2 \left( \frac{1 \, \text{m}}{1 \, \text{G}} \right)
\left( \frac{\partial B}{\partial z} \right)
\label{eq:accel_lin_zeeman} ,
\end{equation}
where $1 \, \text{G} = 10^{-4} \, \text{Tesla}$.
\comment{Similarly, these Zeeman shifts will typically result in $\delta V_n / h \sim 0.2 \, \text{kHz/G}$
and the contribution to $\delta \phi$ of the associated $\delta \mathbf{a}$ is given by}
\begin{equation}
\left( \frac{m}{\Delta m} \right) \delta a \sim 5 \times 10^{4} \,
\text{m/s}^2 \left( \frac{1 \, \text{m}}{1 \, \text{G}} \right) \left( \frac{\partial B}{\partial z} \right)
\label{eq:diff_accel_lin_zeeman} .
\end{equation}
\comment{although $\delta V_n$ and $\delta a$ can be 10 times smaller for particular pairs of initial and final $m_F$ states.}
The systematic effect associated with Eq.~\eqref{eq:diff_accel_lin_zeeman} is anyway rather large unless one considers extremely homogeneous magnetic fields, but it can be effectively suppressed by alternating interferometers that involve atoms with opposite signs of the magnetic quantum number $m_F$. 
\highlight{When doing so, the term proportional to $\delta \mathbf{a} \cdot \bar{\mathbf{a}}$ does not cancel out, but it is seven orders of magnitude smaller than the contribution from Eq.~\eqref{eq:diff_accel_lin_zeeman}.}

Even after cancellation of the linear contributions, those from the \emph{quadratic} Zeeman effect remain and lead to a frequency shift $\delta V_n / h \sim - 0.2 \, \text{Hz} \, (B / 1 \, \text{G})^2$ \cite{boyd07,falke11}, which implies
\begin{equation}
\left( \frac{m}{\Delta m} \right) \, \delta a \sim
10 \, \text{m/s}^2  \left( \frac{B}{0.1 \, \text{G}} \right)
\left( \frac{1 \, \text{m}}{1 \, \text{G}} \right) \left( \frac{\partial B}{\partial z} \right)
\label{eq:diff_accel_quad_zeeman} .
\end{equation}
Hence, reducing this by three orders of magnitude to the $10^{-3} g$ level, requires magnetic field gradients $| \partial B / \partial z | \lesssim 1 \, \text{mG/m}$ for a bias field of $0.1 \, \text{G}$. In fact, thanks to advanced shielding methods, very low gradients $| \partial B / \partial z | \lesssim 2.5 \times 10^{-5} \, \text{G/m}$ have recently been demonstrated for 10-m atomic fountains~\cite{wodey20}, which would bring the contribution from Eq.~\eqref{eq:diff_accel_quad_zeeman} down to the $10^{-5} g$ level.

Such low magnetic gradients would also reduce the values in Eqs.~\eqref{eq:accel_lin_zeeman} and \eqref{eq:diff_accel_lin_zeeman} by more than four orders of magnitude (before any cancellation from combining the results for opposite signs of $m_F$) and would make the term proportional to $\delta \mathbf{a} \cdot \bar{\mathbf{a}}$ completely negligible.


\subsection{Black-body radiation}
\label{sec:bb_radiation}

Black-body radiation at room temperature gives rise to an AC Stark shift of the atomic energy levels proportional to the real part of the electric polarizability and the square of the electric field amplitude, which is in turn proportional to fourth power of the temperature.
Specifically, the two clock states of $^{87}\mathrm{Sr}$ experience a shift of $- 2 \, \text{Hz} \, (T / 300 \, \text{K})^4$ and $- 4 \, \text{Hz} \, (T / 300 \, \text{K})^4$ respectively \cite{porsev08}.
For a uniform (and time-independent) temperature distribution these shifts have no effect on the phase shift $\delta \phi$ because of the internal-state inversion driven by the central pulse. 
Temperature gradients, on the other hand, will lead to state-dependent accelerations that result in a mean acceleration
\begin{equation}
\bar{a} \sim 2 \times 10^{-12}\, \text{m/s}^2
\left( \frac{T}{300 \, \text{K}} \right)^3
\left( \frac{100 \, \text{m}}{1 \, \text{K}} \right) \left( \frac{\partial T}{\partial z} \right)
\label{eq:accel_bbr} .
\end{equation}
More importantly, the relevant contribution from the differential acceleration to $\delta \phi$, which is enhanced by a factor $(m / \Delta m)$, is given by
\begin{equation}
\left( \frac{m}{\Delta m} \right) \delta a \sim 7 \times 10^{-2}\, \text{m/s}^2
\left( \frac{T}{300 \, \text{K}} \right)^3
\left( \frac{100 \, \text{m}}{1 \, \text{K}} \right) \left( \frac{\partial T}{\partial z} \right)
\label{eq:diff_accel_bbr} .
\end{equation}
Thus, for a variation of $2\, \text{K}$ over $100\, \text{m}$, as expected for instance for the facility proposed in Ref.~\cite{arduini23}, temperature gradients would contribute at the $10^{-2} g$ level. By placing temperature sensors along the whole baseline to measure such gradients, their effect can be modeled and post-corrected, which should allow further reduction of the associated systematic uncertainty by at least one order of magnitude down to the $10^{-3}$ level.
\highlight{Furthermore, the effect of temperature inhomogeneities at shorter length scales will be (partially) averaged out when integrating Eq.~\eqref{eq:external_potential_diff} along mid-point trajectories with a total extent comparable to the full baseline.}


\bibliography{literature5}

\end{document}